\pdfoutput=1
\documentclass[aps,prd,floats,floatfix, twocolumn, amsmath,
superscriptaddress,nofootinbib,showpacs,longbibliography]{revtex4-1}
\usepackage[T1]{fontenc}
\usepackage[utf8]{inputenc}
\DeclareUnicodeCharacter{2009}{\nobreakspace}
\usepackage{lmodern}
\usepackage{times}
\usepackage[varg]{txfonts}
\usepackage[normalem]{ulem}
\usepackage{verbatim}
\usepackage[dvipsnames, usenames]{xcolor}
\definecolor{linkcolor}{rgb}{0.0,0.2,0.5}
\usepackage[hypertexnames=false, unicode, colorlinks=true, linkcolor=linkcolor,
citecolor=linkcolor, filecolor=linkcolor,urlcolor=linkcolor,
pdfusetitle]{hyperref}
\usepackage[all]{hypcap}
\usepackage{graphicx}
\usepackage{xspace}
\usepackage{amssymb}
\usepackage{bm} 
\usepackage{microtype}
\usepackage[english]{babel}
\usepackage{aas_macros}
\usepackage{booktabs}

\allowdisplaybreaks

\usepackage{physics}
\usepackage{orcidlink}
\usepackage{multirow}
\usepackage{enumitem}
\DeclareMathAlphabet{\mathpzc}{OT1}{pzc}{m}{it}

\usepackage{titlesec}

\titleclass{\subsubsubsection}{straight}[\subsubsection]
\newcounter{subsubsubsection}[subsubsection]
\renewcommand\thesubsubsubsection{\thesubsubsection.\arabic{subsubsubsection}}
\titleformat{\subsubsubsection}
  {\normalfont\normalsize\bfseries}{\thesubsubsubsection}{1em}{}
\titlespacing*{\subsubsubsection}{0pt}{3.25ex plus 1ex minus .2ex}{1.5ex plus .2ex}

\begin{document}

\preprint{APS/123-QED}

\title{Line-of-Sight Acceleration-driven waveform corrections for unequal-mass compact binary coalescences}

\newcommand{\IUCAA}{\affiliation{Inter-University Centre for Astronomy and
    Astrophysics, Post Bag 4, \\ Ganeshkhind, Pune 411007, India}}
\newcommand{\CITA}{\affiliation{Canadian Institute for Theoretical Astrophysics, University of Toronto, 60 St George St,\\ Toronto, ON M5S 3H8, Canada}}  

\author{Mohd Asim Ansari\,\orcidlink{0009-0000-3217-3844}}
\email{asim.ansari@iucaa.in}
 \IUCAA
\author{Avinash Tiwari\,\orcidlink{0000-0001-7197-8899}}
\email{avinash.tiwari@iucaa.in}
\IUCAA
\author{Sajad A. Bhat\,\orcidlink{0000-0002-6783-1840}}
\IUCAA
\author{Aditya Vijaykumar\,\orcidlink{0000-0002-4103-0666}}
\CITA
\author{Shasvath J. Kapadia\,\orcidlink{0000-0001-5318-1253}}
\email{shasvath.kapadia@iucaa.in}
\IUCAA
\date{\today}

\begin{abstract}
 A constant line-of-sight acceleration (LOSA) of the centre-of-mass of a quasi-circular compact binary coalescence (CBC) leads to modulations in the gravitational-wave (GW) waveform at $-4$ post-Newtonian (PN) order. In our previous work, we calculated the phase and amplitude corrections only for the $(\ell,m) = (2,2)$ mode. The gravitational wave event GW190814 has recently attracted considerable attention because it has been suggested to provide evidence for a nonzero LOSA. However, several LOSA analyses of GW190814 employ higher-order-mode waveform models, such as \texttt{IMRPhenomXPHM}, while incorporating LOSA corrections only for the dominant (2,2) mode. This approximation assumes that their omission has a negligible impact on parameter estimation. In this paper, we present a self-consistent implementation of LOSA in a higher-mode waveform model by deriving the corresponding phase and amplitude corrections. We show that these corrections are essential for asymmetric binaries such as GW190814 and that neglecting them biases parameter inference. We also discuss inconsistencies in the derivation and implementation of LOSA corrections in waveform models in the literature that include higher-order modes.
\end{abstract}

\maketitle


\section{Introduction}
The LIGO-Virgo-KAGRA (LVK) ~\cite{TheLIGOScientific:2014jea, TheVirgo:2014hva, KAGRA:2020tym} catalog of $\sim$300 candidate events released via the second phase of the fourth observing run (O4b) \cite{LIGOScientific:2018mvr,LIGOScientific:2020ibl,KAGRA:2021vkt,LIGOScientific:2025hdt,LIGOScientific:2025yae,LIGOScientific:2025slb,LIGOScientific:2026sit,LIGOScientific:2026ifv,LIGOScientific:2026wfs} offers a broad baseline for compact binary astrophysics. This growing collection of merging binaries provides new windows to explore extragalactic populations and their formation channels \cite{Samsing:2017xmd, Rodriguez:2018pss, Romero-Shaw:2019itr, Samsing:2020tda, Zevin:2021rtf, DallAmico:2023neb}.
Determining the astrophysical origin of compact binary coalescences (CBCs) is one of the primary goals of gravitational-wave (GW) astronomy \cite{Mandel:2021rph}. For most GW  events, however, this remains challenging because electromagnetic counterparts are absent and the sky localization is insufficient to uniquely identify the host galaxy \cite{Singer:2014qca}. Consequently, the formation channel of a CBC is generally inferred statistically from the observed population and the intrinsic properties of the binaries, such as their masses, spins, and orbital eccentricities \cite{Postnov:2014tza, Mapelli:2018, Mandel:2021rph, KAGRA:2021duu, LIGOScientific:2025pvj, LIGOScientific:2020kqk, Farr:2017uvj, Rodriguez:2016vmx}. While these observables provide valuable insight into binary formation, they are generally insufficient to uniquely determine the formation channel of an individual event. 

Previous studies~\cite{Yunes2011,Bonvin2017,Vijaykumar_2023, PhysRevD.110.083008, He:2026bzv, Zhao:2026ydo, Roy:2026mco, Pompili:2026tdf, Roy:2026duh, Gera:2025ugl} have shown that a line-of-sight acceleration (LOSA) of the center of mass (CoM) of a CBC imprints phase and amplitude modulations on the GW signal at $-4$ post-Newtonian (PN) order, providing a potential ``smoking gun'' of the binary's formation channel (see~\cite{Vijaykumar_2023, Yang:2024tje, Hendriks:2026kys, Roy:2026mco, Pompili:2026tdf, Roy:2026duh, Pathak:2026cik, LIGOScientific:2026qni, LIGOScientific:2026fcf} for LOSA constraints from GW events). More recently,~\cite{Tiwari:2024pvb,Tiwari:2025qqx} demonstrated that the LOSA and its higher-order time derivatives\footnote{Note that~\cite{Tiwari:2026lvw, Tiwari:2026cnb} have shown it for a more general case of periodic line-of-sight velocities as well.} encode information about the CoM kinematics, thereby enabling the merger environment — and consequently the binary's formation channel — to be inferred from GW observations. Complementary studies have also investigated environmental effects on GW signals arising from the motion of the binary's CoM~\cite{Santos:2025ass, Hendriks:2026kys, Tagawa:2025tfd, Takatsy:2025bfk, Hendriks:2024gpp, Giri:2026wgy, Camilloni:2023xvf, Cocco:2025adu, Cocco:2025udb}. Collectively, these results establish LOSA as a promising probe of the astrophysical environments and formation channels of individual CBC events.

In~\cite{Vijaykumar_2023}, we derived the LOSA-induced GW phase correction by incorporating nonspinning point-particle contributions up to 3.5PN order. We showed that these higher-order corrections become significant for long-duration signals such as GW170817, whereas they remain negligible for GW170608-like shorter signals when observed with the current-generation (O4) GW detectors. Subsequently,~\cite{PhysRevD.110.083008, Tiwari_pipe_2026} extended the LOSA formalism to aligned-spin CBCs. Furthermore,~\cite{Tiwari_pipe_2026} incorporated tidal effects and demonstrated that several intrinsic physical effects, as well as certain beyond-GR modifications, can produce phase signatures that mimic those arising from LOSA. 

A recent study~\cite{Yang:2024tje} reported evidence for a non-zero LOSA in GW190814~\cite{Abbott_2020_GW190814} and suggested that the binary may have originated in an active galactic nucleus (AGN) disk. Subsequently, \cite{Hendriks:2026kys} argued that this inference resulted from using a relatively short signal duration (4 seconds) during parameter estimation (PE), and showed that extending the signal duration to 32 seconds yields a LOSA consistent with zero. In contrast, in~\cite{Tiwari_pipe_2026}, we proposed an alternative explanation. Owing to the highly asymmetric mass ratio of GW190814, applying a LOSA phase correction derived solely for the dominant $(2,2)$ mode to a waveform containing higher-order modes can introduce a spurious non-zero LOSA even for an intrinsically non-accelerating system. We further argued that this bias can be amplified when phase marginalization~\cite{2019PASA...36...10T} is employed during the PE. In addition, \cite{Pathak:2026cik, Roy:2026mco, Pompili:2026tdf} also analyzed this event and found no support for a non-zero LOSA.

In this paper, we derive the LOSA-induced phase and amplitude corrections for the higher-order modes of the GW waveform. We then perform a series of zero-noise injection and recovery studies for a GW190814-like signal to demonstrate that applying the LOSA corrections derived for the dominant $(2,2)$ mode to all waveform modes biases the inferred CBC parameters, including the LOSA itself. These results establish that LOSA corrections must be applied consistently on a mode-by-mode basis. 

The rest of the paper is organized as follows: In Section ~\ref{sec: methods}, we present the corrections to the waveforms and the method to derive the amplitude corrections. In Section ~\ref{sec: incos}, we discuss the inconsistencies in other similar works, and then present the results in Section~\ref{sec: param_inf}. We conclude in Section~\ref{sec:conclusions}.


\section{Methods}\label{sec: methods}
Let $M$ and $M_{\Gamma}$ denote the redshifted total masses of a non-accelerating and an accelerating CBC, respectively. The observed LOSA of the binary's CoM is parameterized by $\Gamma \equiv a/c$. Hereafter, we work in geometrized units with $G=c=1$. Assuming that $|\Gamma (t_{\rm u}-t_{\rm c})| \ll 1$, the redshifted total mass of the accelerating binary, to first order in $\Gamma$, is given by
\begin{equation}
    \label{eq:redshifted_mass}
    M_{\Gamma}=M\left[1+\Gamma (t_{\rm u}-t_{\rm c})\right],
\end{equation}
where $t_{\rm u}$ denotes the unperturbed time and $t_{\rm c}$ is the coalescence time. 

Under the stationary phase approximation (SPA), the frequency-domain inspiral waveform for the $(\ell,m)$ mode in the presence of LOSA can be written as
\begin{equation}
    \label{eq:WF}
    \tilde{h}_{\ell m,\Gamma}(f)
    =
    \left[
    1+\frac{\Delta \mathcal{A}_{\ell m,\Gamma}}{\mathcal{A}_{\ell m}}(f)
    \right]
    e^{-i\Delta\Psi_{\ell m,\Gamma}(f)}
    \tilde{h}_{\ell m}(f),
\end{equation}
where $\tilde{h}_{\ell m}(f)$ denotes the unperturbed waveform for the $(\ell,m)$ mode~\cite{Pratten:2020ceb, Garcia-Quiros:2020qpx}. To leading order in $\Gamma$, the LOSA-induced phase and amplitude corrections are given by
\begin{equation}
    \label{eq:phas_corr}
    \Delta \Psi_{\ell m,\Gamma}(f)
    =
    \frac{m}{2}\,
    \frac{25\,\Gamma M}
    {65536\,\eta^2\,v_m^{13}},
\end{equation}
and
\begin{equation}
    \label{eq:amp_cor}
    \frac{\Delta \mathcal{A}_{\ell m,\Gamma}}
    {\mathcal{A}_{\ell m}}(f)
    =
    \zeta_{\ell m}
    \frac{15\,\Gamma M}
    {512\,\eta\,v_m^8},
\end{equation}
respectively. Here,
\begin{equation}
    \label{eq:vm_def}
    v_m \equiv \left(\frac{2\pi M f}{m}\right)^{1/3},
\end{equation}
where $f$ is the GW frequency and $\eta$ is the symmetric mass ratio. The coefficient $\zeta_{\ell m}$ is a mode-dependent rescaling factor, with $\zeta_{22}=1$. For the subdominant modes, their values are
\begin{equation}
    \zeta_{21}=\zeta_{33}=\frac{7}{9},
    \qquad
    \zeta_{32}=\zeta_{44}=\frac{5}{9}.
\end{equation}

The LOSA-induced phase correction, including higher-order nonspinning point-particle contributions, is given by $(m/2)\Delta\Psi(2f/m)$, while the corresponding correction including aligned-spin effects is $(m/2)\left[\Delta\Psi(2f/m)+\Delta\Psi_{\rm AS}(2f/m)\right]$. Here, $\Delta\Psi(f)$ is given by Eq.~(4) of  ~\cite{Vijaykumar_2023}, and $\Delta\Psi_{\rm AS}(f)$ by Eq.~(13) of  ~\cite{Tiwari_pipe_2026}; in both expressions, the frequency $f$ is replaced by $2f/m$.

For parameter estimation (PE), we modify the \texttt{IMRPhenomXPHM}~\cite{Colleoni:2023czp} waveform model to incorporate the LOSA effect. Specifically, we ({\it i}) generate the frequency-domain waveform for each $(\ell,m)$ mode, $\tilde{h}_{\ell m}$, separately; ({\it ii}) apply the LOSA-induced phase and amplitude corrections using Eq.~(\ref{eq:WF}) to obtain $\tilde{h}_{\ell m,\Gamma}$; and ({\it iii}) sum the modes to construct the final waveform.

\subsection{Derivation of the Amplitude Correction}
In this section, we derive the amplitude correction to the $(\ell,\,m)$ mode due to LOSA. Specifically, we use the $dE/f$ {\it invariance} to compute the amplitude corrections, where $E$ is the energy carried by GWs and $f$ is the GW frequency. Below we delineate the method. 

The energy spectrum of the GWs, for the (2, 2) mode, is defined as (see Equation~(1.160) of~\cite{maggiore2007gravitational})
\begin{equation}
    \label{eq: energy_spec}
    \frac{dE}{df} = \frac{\pi c^3}{2 G} f^2 D_{\rm L}^2 \int d\Omega \left( \vert \tilde{h}_+ (f) \vert^2 + \vert \tilde{h}_\times (f) \vert^2 \right),
\end{equation}
where $d\Omega$ is the differential solid angle. We can rewrite this equation as
\begin{equation}
    \label{eq: dE_by_f}
    \frac{dE}{f} = \frac{\pi c^3}{2 G} D_{\rm L}^2 f df \int d\Omega \left( \vert \tilde{h}_+ (f) \vert^2 + \vert \tilde{h}_\times (f) \vert^2 \right)
\end{equation}
Notice that the quantity $dE/f$ is invariant when GWs undergo any kind of redshifting because $dE \to dE / (1 + z(t))$ and $f \to f / (1 + z(t))$, where $z(t)$ is the redshift, and hence it can be used to compute the frequency-domain amplitude corrections due to effects like the Doppler shift. 

Since $\left| \tilde{h}_{+}(f) \right|^{2} + \left| \tilde{h}_{\times}(f) \right|^{2} = \left| \tilde{h}_{+} - i \tilde{h}_{\times} \right|^2$ and $ \tilde{h}_{+} - i \tilde{h}_{\times} = \sum_{\rm \ell, m} \tilde{h}_{\rm \ell m} \mathbf{Y}_{-2}^{\rm \ell m}(\theta, \phi) $, we can re-write Equation~\eqref{eq: dE_by_f} for a GW signal having higher modes as
\begin{equation}
    \frac{dE}{df} = \frac{\pi c^{3}}{2G} f^{2} D_{\rm L}^{2} \int d\Omega \left|\sum_{\rm \ell, m} \tilde{h}_{\rm \ell m} \mathbf{Y}_{-2}^{\rm \ell m}(\theta, \phi) \right|^2  
    \label{eq: dE_by_df_hm_}
\end{equation}
Due to the orthogonality of $\mathbf{Y}_{-2}^{\rm \ell m}$, the interference terms between different modes will integrate to zero. Therefore, we can further write
\begin{equation}
    \frac{dE}{df} \propto f^{2} \sum_{\rm \ell, m}\left|\tilde{h}_{\rm \ell m}\right|^2
    \label{eq: dE_by_df_hm}
\end{equation}
or equivalently
\begin{equation}
    \frac{dE}{f} \propto f df \sum_{\rm \ell, m}\left|\tilde{h}_{\rm \ell m}\right|^2 
    \label{eq: dE_by_f_hm}
\end{equation}
Because each spherical harmonic mode chirps at a different frequency, $dE/f$ will be invariant for each mode separately. Therefore, we can write
\begin{equation}
    \left( \frac{dE}{f} \right)_{\ell m} \propto f df \left|\tilde{h}_{\rm \ell m}\right|^2 = f df \mathcal{A}_{\ell m}^2(f).
    \label{eq: dE_by_f_lm}
\end{equation}
Observing that, to the linear order in $\Gamma$,
\[ df_{\rm u} = df_{\rm o} (1 + \Gamma (t_{\rm u} - t_{\rm c})) \left[1 + \frac{f_{\rm o} \Gamma}{df_{\rm o}/dt_{\rm o}} \right], \]
using the invariance of $(dE/f)_{\ell m}$, we can write
\begin{equation}
    \mathcal{A}_{\ell m, \rm o} (f_{\rm o}) = \mathcal{A}_{\ell m, \rm u} (f_{\rm u}) \left(1 + \Gamma (t_{\rm u} - t_{\rm c}) \right) \left[1 + \frac{f_{\rm o} \Gamma}{df_{\rm o}/dt_{\rm o}} \right]^{1/2}
    \label{eq: A_lm_trans_t}
\end{equation}
Using Equations~(A1)- (A4) of~\cite{Vijaykumar_2023} with $v_{m, \rm u}$ and $v_{m, \rm o}$ in places of $v_{\rm u}$ and $v_{\rm o}$, respectively, we can write
\begin{align}
    \label{eq: dv_dt_red}
    \frac{dv_{m, \rm o}}{dt_{\rm o}} &= \frac{32 \eta v_{m, \rm o}^9}{5 M} \left(1 - \frac{65}{768}\frac{M \Gamma}{\eta v_{m, \rm o}^8} \right) \\
    1 + \Gamma (t_{\rm u} - t_{\rm c}) &= 1 - \frac{5 M \Gamma}{256 \eta v_{m,\rm o}^8}
\end{align}

It can straightforwardly be seen that $\dot{f}_{\rm o} / f_{\rm o} = 3 \dot{v}_{m,\rm o} / v_{m, \rm o}$. Therefore using Equations~\eqref{eq: dv_dt_red} in combination with $\dot{f}_{\rm o} / f_{\rm o}$ substitution in Equation~\eqref{eq: A_lm_trans_t}, we can write
\begin{equation}
    \mathcal{A}_{\ell m, \rm o} (f_{\rm o}) = \mathcal{A}_{\ell m, \rm u} (f_{\rm u}) \left( 1 + \frac{5 M \Gamma}{768 \eta v_{m,\rm o}^8} \right)
    \label{eq: A_lm_trans_vm}
\end{equation}

To leading order, $\mathcal{A}_{\ell m, \rm u}$ be $\propto f_{\rm u}^{-\alpha}$. We can then write 
\begin{equation}
    \mathcal{A}_{\ell m, \rm o} (f_{\rm o}) \propto f_{\rm o}^{-\alpha} \left( 1 + \frac{5 (3 \alpha + 1) M \Gamma}{768 \eta v_{m,\rm o}^8} \right).
    \label{eq: A_lm_trans_vm_o}
\end{equation}
We can then write the leading-order relative amplitude correction as
\begin{equation}
    \frac{\Delta \mathcal{A}_{\ell m, \Gamma}}{\mathcal{A}_{\ell m}} = \frac{5 (3 \alpha + 1) M \Gamma}{768 \eta v_{m}^8},
    \label{eq: A_lm_trans_vm_o}
\end{equation}
where we have replaced $v_{m,\rm o}$ by $v_m$ as we have already assumed the SPA and $\zeta_{\ell m}$ in Equation~\eqref{eq:amp_cor} is related to $\alpha$ by $\zeta_{\ell m} = 2 (3 \alpha + 1) / 9$. 

For the $(2,2)$ mode, $\alpha = 7/6$, i.e., $\mathcal{A}_{22} \propto f^{-7/6}$. Therefore, we can write
\begin{equation}
    \frac{\Delta\mathcal{A}_{22,\Gamma}}{\mathcal{A}_{22}} = \frac{15}{512}\frac{M \Gamma}{\eta v_2^8}.
    \label{eq:rel_amp_err_22}
\end{equation}
Note that this is different from the amplitude correction derived in \cite{Vijaykumar_2023}. This is because \cite{Vijaykumar_2023} assumed $A(t)$ to remain invariant and the dependence of $A(t)$ on $(t_{\rm c} - t)$ to remain the same, and substituted the new stationary time, which is incorrect because while $A(t)$ remains invariant, it is $dt$ which transforms due to the time dilation, not $t$, and hence $A(t)$ will not have the same dependence on $(t_{\rm c} - t)$.

Because for $(2,1)$ and $(3,3)$ modes, $\alpha = 5/6$, the relative amplitude corrections become
\begin{align}
    \frac{\Delta\mathcal{A}_{21,\Gamma}}{\mathcal{A}_{21}} &= \frac{35}{1536}\frac{M \Gamma}{\eta v_1^8} \\
    \frac{\Delta\mathcal{A}_{33,\Gamma}}{\mathcal{A}_{33}} &= \frac{35}{1536}\frac{M \Gamma}{\eta v_3^8}.
    \label{eq:rel_amp_err_21_33}
\end{align}

Similarly, for $(3,2)$ and $(4,4)$ modes, in which case $\alpha = 1/2$, the relative amplitude corrections become
\begin{align}
    \frac{\Delta\mathcal{A}_{32,\Gamma}}{\mathcal{A}_{32}} &= \frac{25}{1536}\frac{M \Gamma}{\eta v_2^8} \\
    \frac{\Delta\mathcal{A}_{44,\Gamma}}{\mathcal{A}_{44}} &= \frac{25}{1536}\frac{M \Gamma}{\eta v_4^8}.
    \label{eq:rel_amp_err_32_44}
\end{align}
Note that these corrections match those in \cite{Roy:2026mco}, which uses a different method.

\section{Inconsistencies in Other Similar Works}\label{sec: incos} 

In a recent study, \cite{Roy:2026mco} computed the phase and amplitude corrections for higher harmonics of the GWs. Specifically, they used the Rømer delay interpretation to compute the time delay and then computed the phase correction to the GW waveform given by $\Delta \Psi (f) = - 2 \pi f \Delta t (f)$. However, they did not consider the Doppler shift in the GW frequency and did not account for the change in the stationary time while using the SPA. These omissions lead to mass-redshift degeneracy breaking even in the case of a constant redshift, as well as a violation of the SPA. We delineate this below.

After computing $\Delta t$ in Eq. (6), which is given by $\Delta t \equiv t_{\rm det} - t_{\rm src} = z_0 t_{\rm src} + (1/2) \Gamma t_{\rm src}^2$, Roy et al.~\cite{Roy:2026mco} claim that the term proportional to $z_0$, corresponding to a constant Doppler shift, can be absorbed in the redefinition of the masses and the coalescence time. Setting $\Gamma = 0$, one gets $t_{\rm det} = (1 + z_0) t_{\rm src}$, which is the expected time dilation. However, once we substitute $t_{\rm src} (f_{\rm src}) = - 5 M / (256 \eta v_{\rm src}^8)$, where $v_{\rm src} = (\pi M f_{\rm src})^{1/3}$, we get $t_{\rm det} = - 5 M (1 + z_0)/ (256 \eta v_{\rm src}^8)$. Notice that the numerator of $t_{\rm det}$ has the term $M (1 + z_0)$, while the denominator still has only $M$. Unless one also Doppler-shifts the frequency, one would not recover the mass-redshift degeneracy, because it requires that $t_{\rm det}$ have the same functional form as $t_{\rm src}$, with $M \to M(1 + z_0)$. Therefore, one must also Doppler shift the GW frequency. Since Doppler shifting of the GW frequency is required for the constant Doppler-shift scenario, it is straightforward to see that the same would be required when the redshift is time-varying. 

If one assumes that $f_{\rm src}$ is the observed frequency, the SPA will require that the stationary time $t_\star = (1/2\pi) d (\Psi + \Delta \Psi)/df_{\rm src}$, which gives
\begin{equation}
    \label{eq: time_soumen}
    t - t_{\rm c} = - \frac{5 M}{256 \eta v_{\rm src}^8} \left[1 + \frac{65 \Gamma M}{1536 \eta v_{\rm src}^8}  \right].
\end{equation}
Using Eq. (7) of  ~\cite{Roy:2026mco}, we can write
\begin{equation}
    \label{eq: tdet_soumen}
    t_{\rm det} - t_{\rm c} = t_{\rm src} + \Delta t (f_{\rm src}) = - \frac{5 M}{256 \eta v_{\rm src}^8} \left[1 + \frac{5 \Gamma M}{512 \eta v_{\rm src}^8}  \right],
\end{equation}
which is different from Eq.~\eqref{eq: time_soumen}. Therefore, the SPA is also not satisfied if only time delay is considered without Doppler shifting the GW frequency.

We now delineate why \cite{Roy:2026mco} still gets the correct phase correction even though they employ an inconsistent method. The total SPA phase is given by 
\begin{equation}
    \label{eq:total_phase}
    \Psi_{\rm tot} = 2 \pi f t_\star - 2 \phi (t_\star) - \frac{\pi}{4} + \pi \Gamma f (t_\star- t_{\rm c})^2,
\end{equation}
where $f$ is the Fourier frequency, which will be equal to the observed GW frequency. However,  ~\cite{Roy:2026mco} assumes $t_\star = t_{\rm src} (f)$, i.e., it sets $f = f_{\rm src}$. Using Newtonian order expressions $t_{\rm src} = t_{\rm c} -\frac{5 M }{256 \eta v_{\rm src}^8}$ and $\phi = \phi_{\rm c} -\frac{1}{32 \eta v_{\rm src}^5}$ along with $f = \frac{v^3}{\pi M}$ in Eq. ~\eqref{eq:total_phase}, we get
\begin{multline}
    \Psi_{\rm tot} = \frac{2v^3}{M} \left(t_{\rm c} - \frac{5 M }{256 \eta v_{\rm src}^8}\right) - 2 \left(\phi_{\rm c} - \frac{1}{32 \eta v_{\rm src}^5} \right) \\ -\frac{\pi}{4} + \frac{\Gamma v^3}{M} \left(-\frac{5 M }{256 \eta v_{\rm src}^8}\right)^2
\end{multline}
Since $f$ will be equal to the observed GW frequency $f_{\rm o}$, we can write $f_{\rm src} = f \left(1 + \Gamma (t_{\rm u} - t_{\rm c}) \right)$ ---   ~\cite{Roy:2026mco} also realizes this while computing the amplitude correction, while it does not do the same for the phase correction. If we make this substitution in the above equation, using the fact that, to the leading order in $\Gamma$, $\Gamma t_{\rm u} = \Gamma t_{\rm o}$, the additional contributions of $\Gamma$ from the first two terms cancel each other, and we get
\begin{equation}
    \label{eq:cor_phase_cor}
    \Psi_{\rm tot} = 2 \pi f t_{\rm c} - 2 \phi_{\rm c} -\frac{\pi}{4} + \frac{3}{128 \eta v^5} + \frac{25 M \Gamma}{65536 \eta^2 v^{13}}
\end{equation}
which matches the literature~\cite{Bonvin2017, Vijaykumar_2023} and is what \cite{Roy:2026mco} also gets. However, in the case of \textit{eccentric} CBCs, these terms do not cancel each other, and we get an additional term even in the leading-order phase correction~\cite{Ansari:Ecc_2026}. We refer the reader to  ~\cite{Ansari:Ecc_2026} for further details on the eccentric case.

\begin{table}[h]
    \centering
    \begin{tabular}{|c|c|}
        \hline 
        \textbf{Parameter} & \textbf{Prior} \\
        \hline \hline
        LOSA $\Gamma [{\rm s}^{-1}]$   & $\mathcal{U}$(-0.1, 0.1)   \\
        \hline
        Chirp Mass $\mathcal{M} [\rm M_{\odot}]$   & $\mathcal{U}$(5.5, 6.75) \\
        \hline
        Mass ratio $q$   & $\mathcal{U}$(0.01, 0.5)  \\
        \hline
        Luminosity Distance $D_{\rm L} [\, {\rm Mpc}]$  & PL(10, 2000) $\propto D_{\rm L}^2$ \\
        \hline
        $\chi_1$ &  $\mathcal{U}$ (0, 0.5) \\
        \hline
        $\chi_2$ &  $\mathcal{U}$ (0, 0.5) \\
        \hline 
        $\rm RA$ & $\mathcal{U}$(0, $2\pi$) \\
        \hline
        $\rm DEC$ & Cosine($-\pi / 2$, $\pi / 2)$ \\
        \hline
        $\theta_{\rm JN}$ & Sine(0, $\pi$) \\
        \hline
        Polarization angle $\psi$ & $\mathcal{U}$(0, $\pi$) \\
        \hline
        Phase at the reference frequency $\phi$ & $\mathcal{U}$(0, $2\pi$) \\
        \hline
        Geocent time $t_c$ & $\mathcal{U}(-0.2, 0.2)$ \\
        \hline
    \end{tabular}
    \caption{The table of priors used during PE of all injections: $\mathcal{U}$ refers to uniform priors while PL refers to power law. Note that we fix $\chi_{1,2}$ to zero and the prior on $t_c$ is around the GPS time (1249852257.0).}
    \label{tab:priors}
\end{table}

\section{Results}\label{sec: param_inf}

\subsection{Testing the Implementation}\label{sec: param_inf_test}
Following the prescription in Section~\ref{sec: methods}, we inject a GW190814-like aligned-spin, non-accelerating system in zero-noise. We set the injection parameters to: $\{ \mathcal{M} = 6.11 \, {\rm M}_{\odot},\, q = 0.1116,\, D_{\rm L} = 230 \, {\rm Mpc},\, t_{\rm c} = {\rm GPS \, time}\, (1249852257.0),\, \chi_1 = 0.07,\, \chi_2 = 0.02,\, {\rm RA} = 0.24,\, {\rm DEC} = -0.6,\, \theta_{\rm JN} = 0.9,\, \psi = 0.54,\, \phi = 2.95,\, \Gamma = 0\}$. Here the mass ratio is defined as the mass of a lighter object divided by the mass of a heavier object. We then estimate the parameters following the same prescription and using the priors listed in Table~\ref{tab:priors} and the \textsc{Dynesty}~\cite{2020MNRAS.493.3132S} sampler implemented in \textsc{Bilby}~\cite{bilby_paper}. Note that throughout the analyses in this {\it paper}, we fix the signal duration to 16 seconds unless specified otherwise, assume the LVK network to be operating at O5 sensitivity\footnote{We use the PSDs available at \href{https://dcc.ligo.org/LIGO-T2000012-v2/public}{https://dcc.ligo.org/LIGO-T2000012-v2/public}.}, and compute the likelihood using the standard \texttt{GravitationalWaveTransient} likelihood for CBCs over the frequency range $20$--$1792~\mathrm{Hz}$ with a sampling rate of $4096~\mathrm{Hz}$. In addition, we use the priors tabulated in~\ref{tab:priors} for all analyses in this paper, except for $\mathcal{M}$ and $q$, for which we explicitly state whenever a different prior is used.

\begin{figure}
    \centering
    \includegraphics[width=0.975\linewidth]{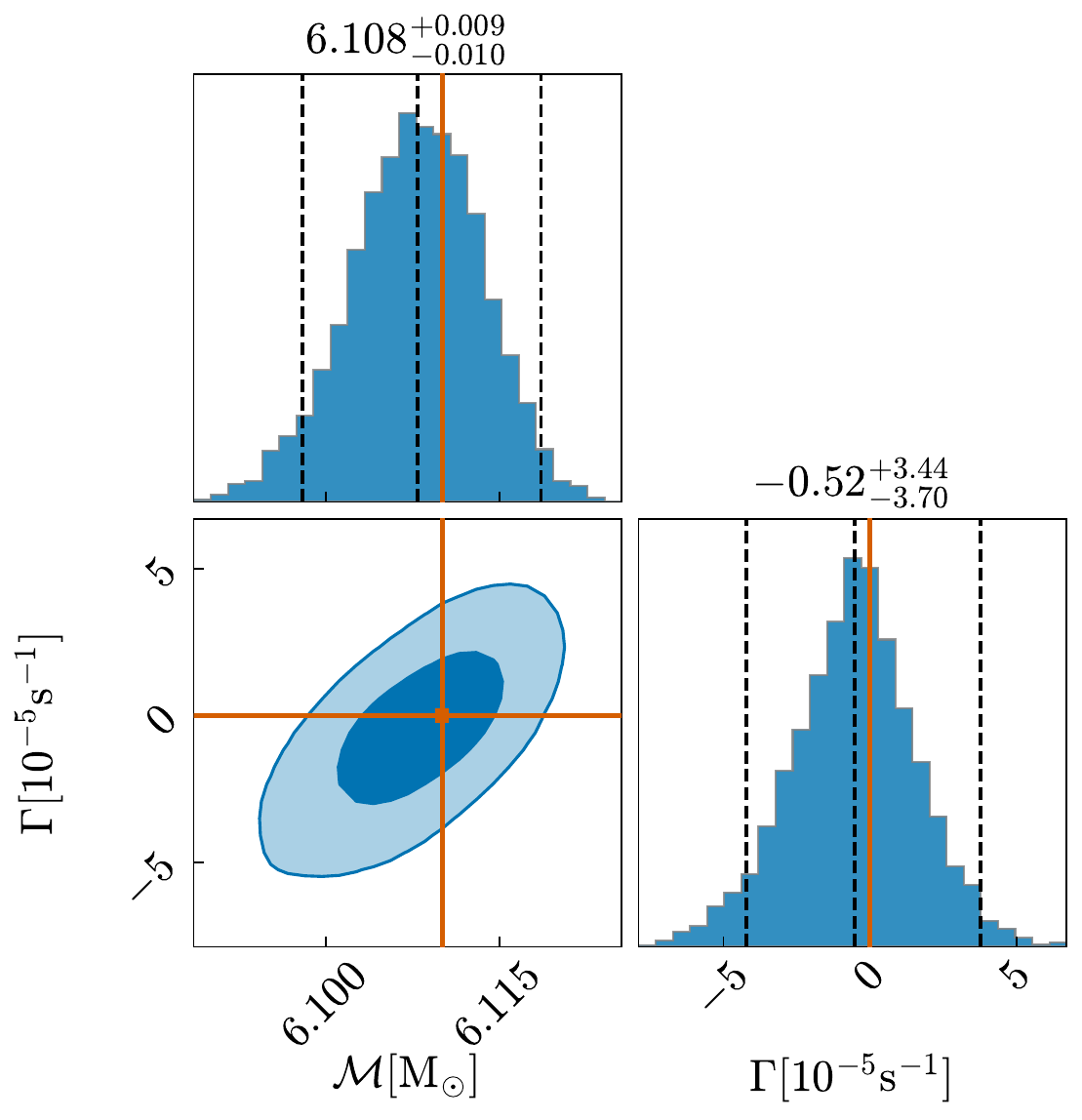}
    \caption{Posteriors of $\mathcal{M}$ and $\Gamma$ for a GW190814-like non-accelerating zero-noise injection, when recovered with a waveform with higher modes and correctly applied LOSA phase and amplitude correction.}
    \label{fig: correct_0losa_rec}
\end{figure}

Figure~\ref{fig: correct_0losa_rec} shows the posteriors of $\mathcal{M}$ and $\Gamma$ for the same. We validate that all parameters are recovered within the 90\% credible interval (CI).

\subsection{Impact of incorrectly applying LOSA modulations to all modes}\label{sec: all_modes_bias}
Once we have tested our prescription for a non-accelerating system, we show biases in the estimates of the parameters as a result of incorrect application of the (2,2) mode phase and amplitude corrections due to LOSA onto all modes for an accelerating system, as was previously done in \cite{Yang:2024tje}. For this purpose, we inject the same GW190814-like system as in Section~\ref{sec: param_inf_test} with finite LOSA following the prescription in Section~\ref{sec: methods} and recover the parameters using a waveform where we add the (2,2) mode phase and amplitude corrections due to LOSA onto \textsc{IMRPhenomXPHM}. Specifically, we consider two cases $\Gamma = - 5 \times 10^{-5} {\rm s^{-1}}$ and $\Gamma = 5 \times 10^{-5} {\rm s^{-1}}$, ensuring that $\vert \Gamma t_{\rm sd} \vert \approx 0.00054 \ll 1$ is valid and the special relativistic effects can be neglected. Here $t_{\rm sd}$ is the signal duration and has been calculated using \cite{maggiore2007gravitational} 
\begin{equation}
    t_{\rm sd} = \frac{5}{256} \mathcal{M}^{-5/3}(\pi f_{\rm min})^{-8/3}.
    \label{eq: tsd}
\end{equation} 
For a GW190814-like system, it comes out to be $10.75 \, {\rm s}$ assuming the minimum frequency $f_{\rm min} = 20 {\rm Hz}$.  

\begin{figure}[ht!]
    \centering
    \includegraphics[width=0.975\linewidth]{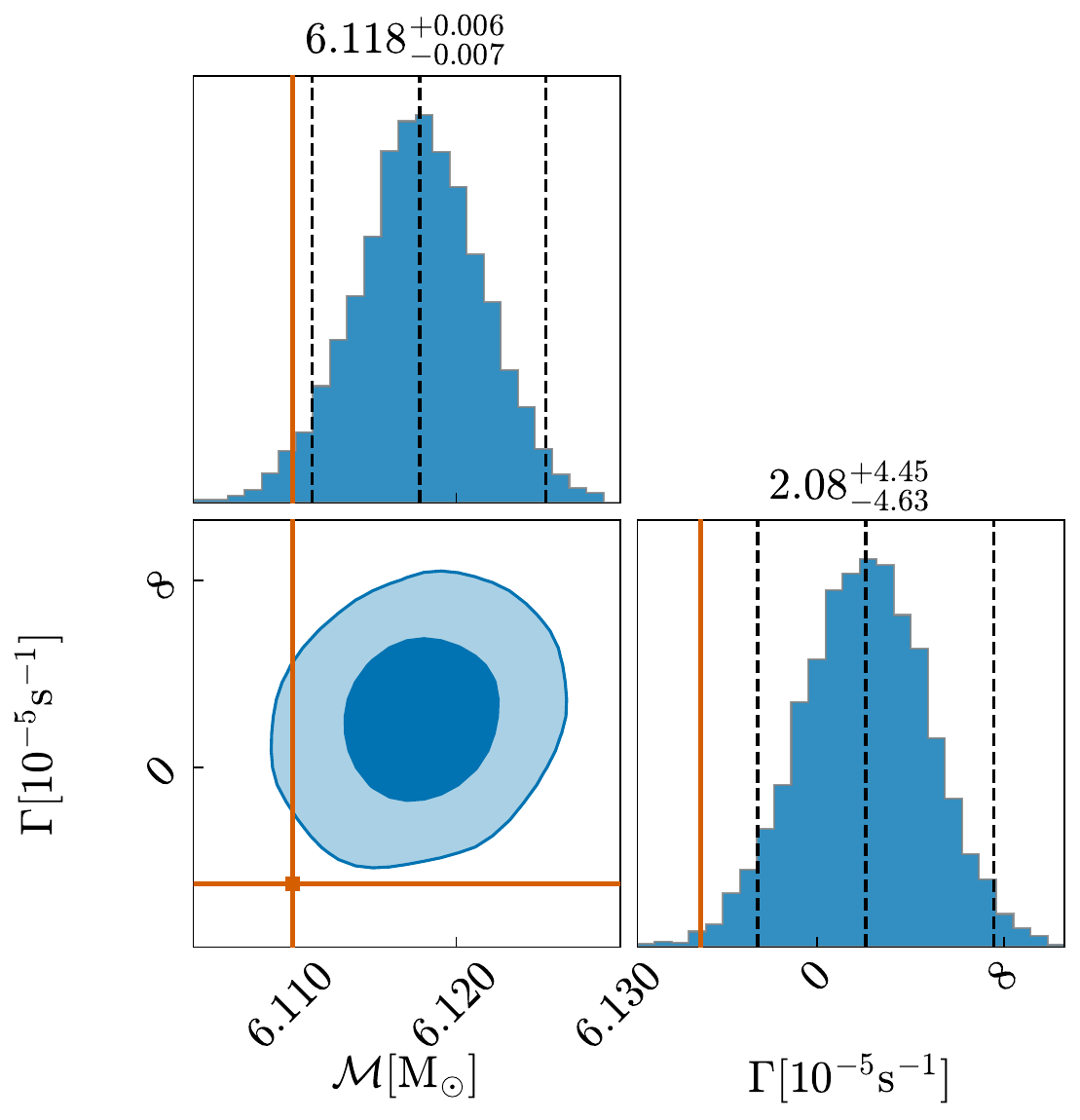}
    \caption{Posteriors of $\mathcal{M}$ and $\Gamma$ for a GW190814-like zero-noise injection, when recovered with a waveform with higher modes and incorrectly applied LOSA phase and amplitude correction of the $(l,m) = (2,2)$ mode to all modes for $\Gamma = -5 \times 10^{-5}{\rm s^{-1}} $ injection.}
    \label{fig: bias_LOSA}
\end{figure}

Figure~\ref{fig: bias_LOSA} shows posteriors of $\mathcal{M}$ and $\Gamma$ for the negative LOSA injection, while Figure~\ref{fig: bias_pos_LOSA} in Appendix~\ref{app: add_plots} shows the same for the positive LOSA case. We find that neither $\mathcal{M}$ nor $\Gamma$ is measured within the 90\% CI for either of the two cases.

\begin{figure}[h!]
    \centering
    \includegraphics[width=0.975\linewidth]{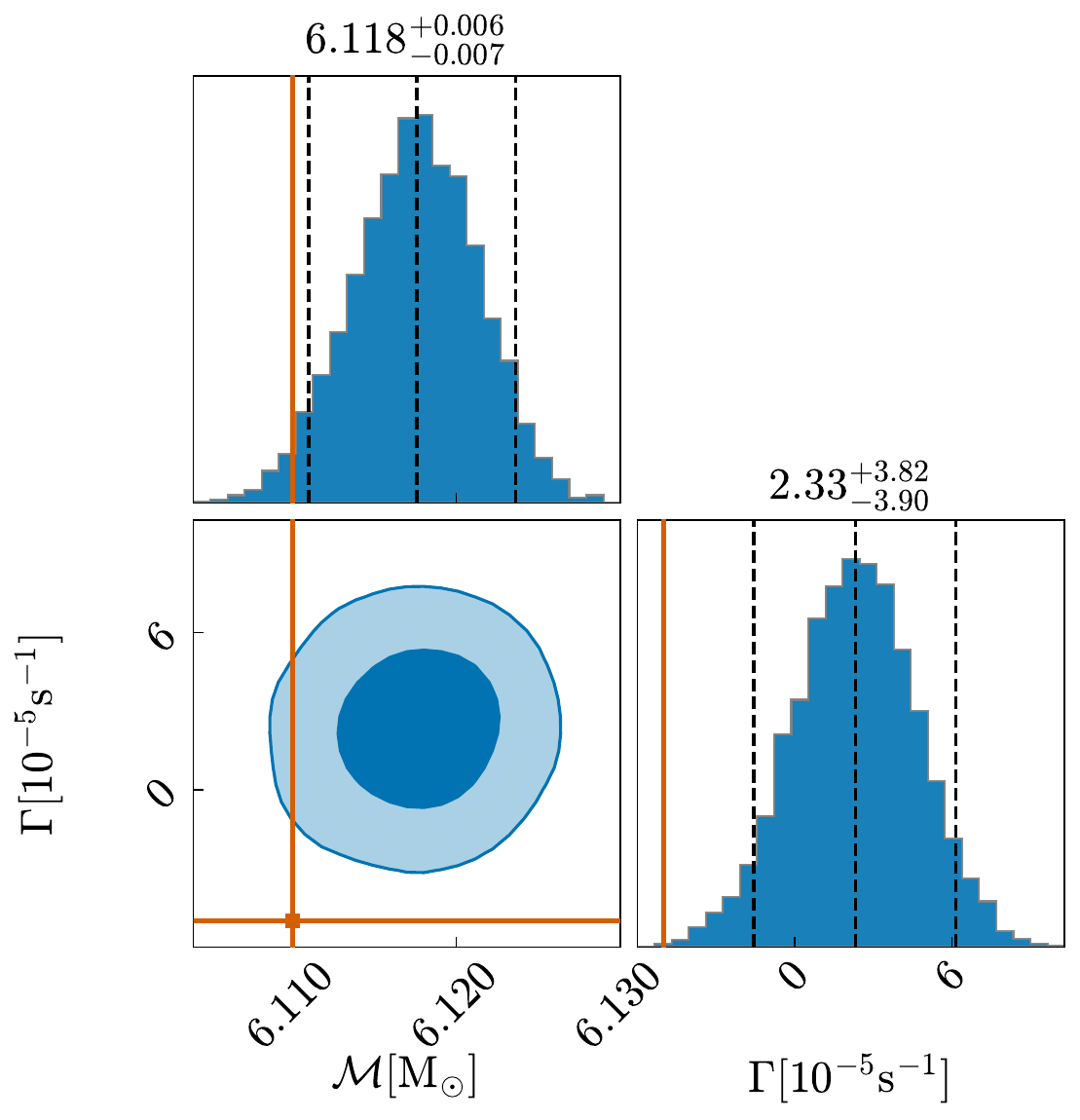}
    \caption{Posteriors of $\mathcal{M}$ and $\Gamma$ for a GW190814-like zero-noise negative LOSA injection, when recovered with a waveform with higher modes and an incorrectly applied LOSA phase and amplitude correction of the $(l,m) = (2,2)$ mode to only the (2,2) mode, keeping the subdominant modes unaffected.}
    \label{fig: bias_LOSA_v2}
\end{figure}

\subsection{Bias Due to LOSA Corrections Applied Only to the (2,2) Mode }\label{sec: only_22_mode_bias}
Another incorrect way to apply the LOSA correction is to modulate only the dominant (2, 2) mode while the subdominant modes remain unaffected. Here we show that neglecting LOSA corrections to the subdominant mode will lead to a biased parameter recovery. For the injection setup, we follow the same prescription as in Section~\ref{sec: methods} and~\ref{sec: param_inf_test}, including the two cases $\Gamma = 5 \times 10^{-5}s^{-1}$ and $- 5 \times 10^{-5}s^{-1}$. The posteriors on $\mathcal{M}$ and $\Gamma$ for negative LOSA have been shown in Figure~\ref{fig: bias_LOSA_v2}, while the same for positive LOSA have been shown in Figure~\ref{fig: bias_pos_LOSA_v2} in Appendix~\ref{app: add_plots}. In both cases, we observe a biased recovery of $\mathcal{M}$ and $\Gamma$.

\subsection{Ineffectiveness of Amplitude Corrections}
\label{sec: no_amp_cor}
\begin{figure}[ht!]
    \centering
    \includegraphics[width=0.975\linewidth]{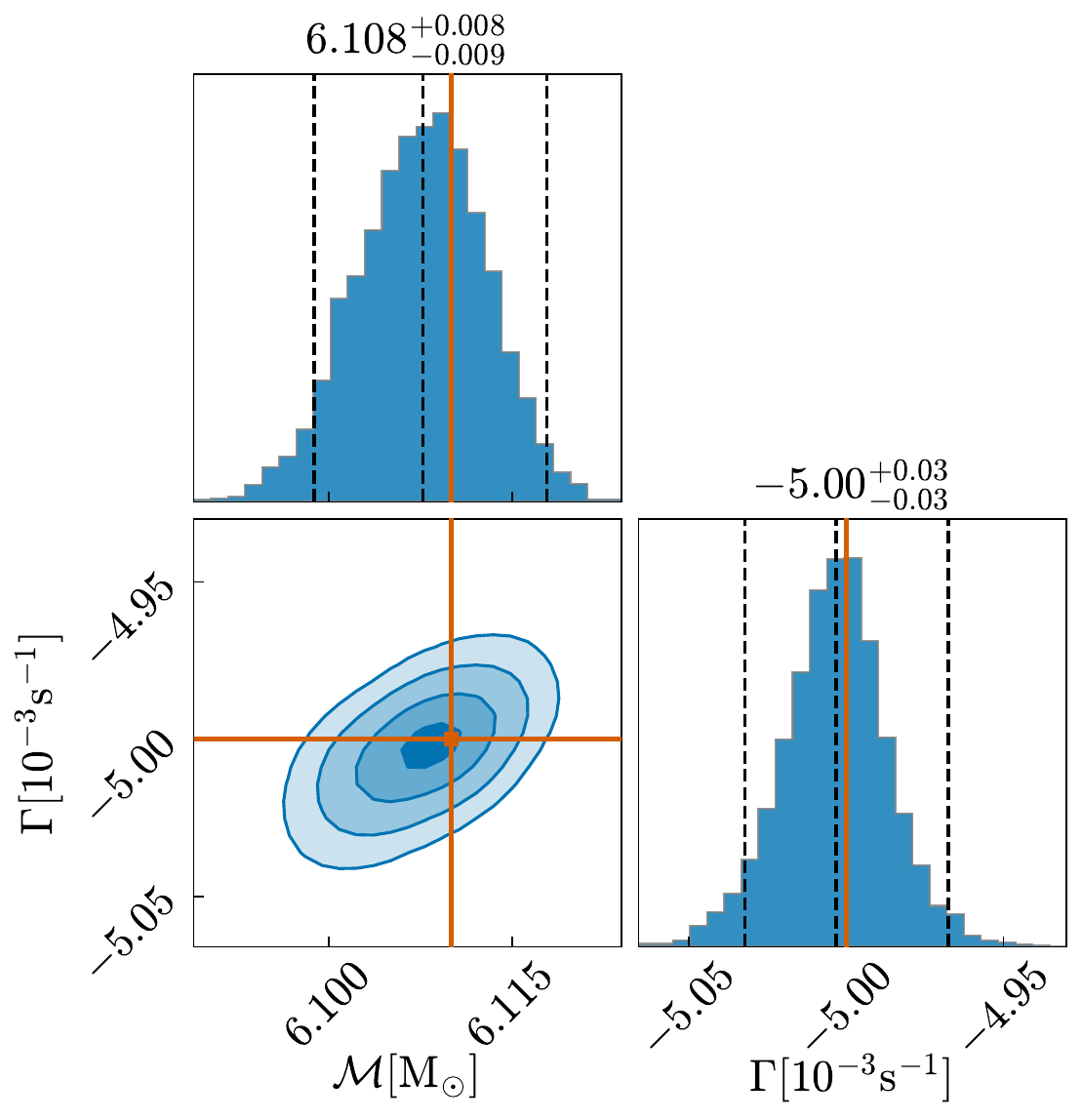}
    \caption{Posteriors of $\mathcal{M}$ and $\Gamma$ for a GW190814-like zero-noise negative LOSA injection when recovered with a waveform without amplitude corrections.}
    \label{fig: amp_corr}
\end{figure}

We also carry out a bias study pertaining to the impact of amplitude corrections on parameter estimation. For this case, we neglect the amplitude corrections and keep the corrections only to the phases of the modes in the inference waveform model. We performed the PE runs following the same injection settings as in Section~\ref{sec: param_inf_test} for two injected LOSA values: $\Gamma = - 5 \times 10^{-3} {\rm s^{-1}}$ and $\Gamma = 5 \times 10^{-3} {\rm s^{-1}}$. Figure~\ref{fig: amp_corr} shows the posteriors on $\mathcal{M}$ and $\Gamma$ for negative LOSA, while Figure~\ref{fig: amp_corr_v2} in Appendix~\ref{app: add_plots} shows the same for positive LOSA. Unlike the previous cases, here we keep LOSA high so that the biases can be captured easily. However, we find that ignoring the amplitude corrections does not produce a noticeable bias in any of the CBC parameters. 

\subsection{Self-consistency Test}
\label{sec: self_cons}
\begin{figure}[ht!]
    \centering
    \includegraphics[width=0.975\linewidth]{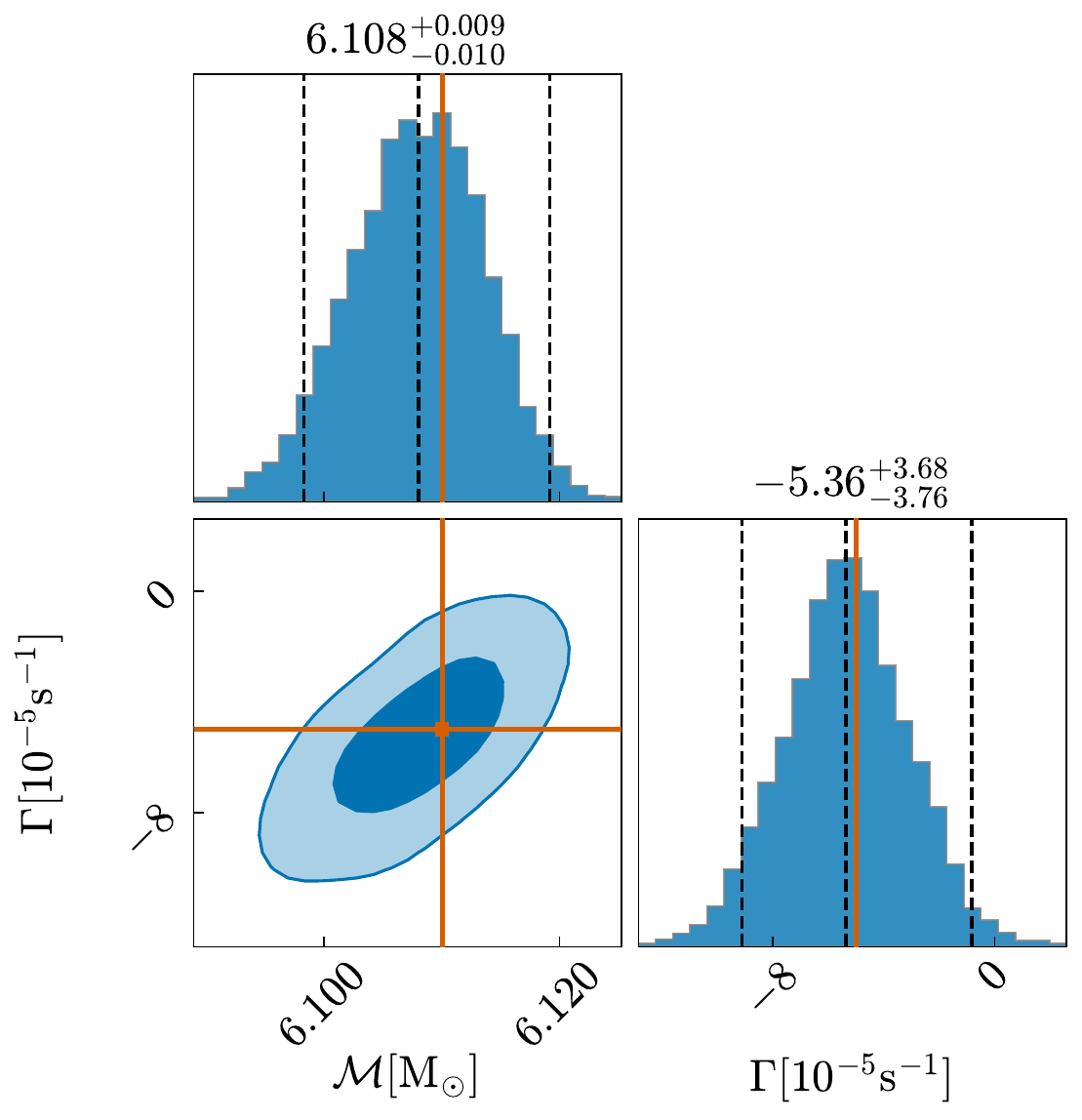}
    \caption{Posteriors of $\mathcal{M}$ and $\Gamma$ for a GW190814-like zero-noise negative LOSA injection, when recovered with a waveform with higher modes and correctly applied LOSA phase and amplitude correction.}
    \label{fig: self_consist}
\end{figure}

We now perform the PE following the prescription in Section~\ref{sec: methods} for computing the waveforms. Figure~\ref{fig: self_consist} shows the posteriors of $\mathcal{M}$ and $\Gamma$ for the negative LOSA injection; Figure~\ref{fig: self_consist_v2} shows the same for the positive LOSA injection. Notice that both parameters $\mathcal{M}$ and $\Gamma$ are measured within the 90\% CI for the cases. 

\begin{figure*}[ht!]
    \centering
    \includegraphics[width=0.485\linewidth]{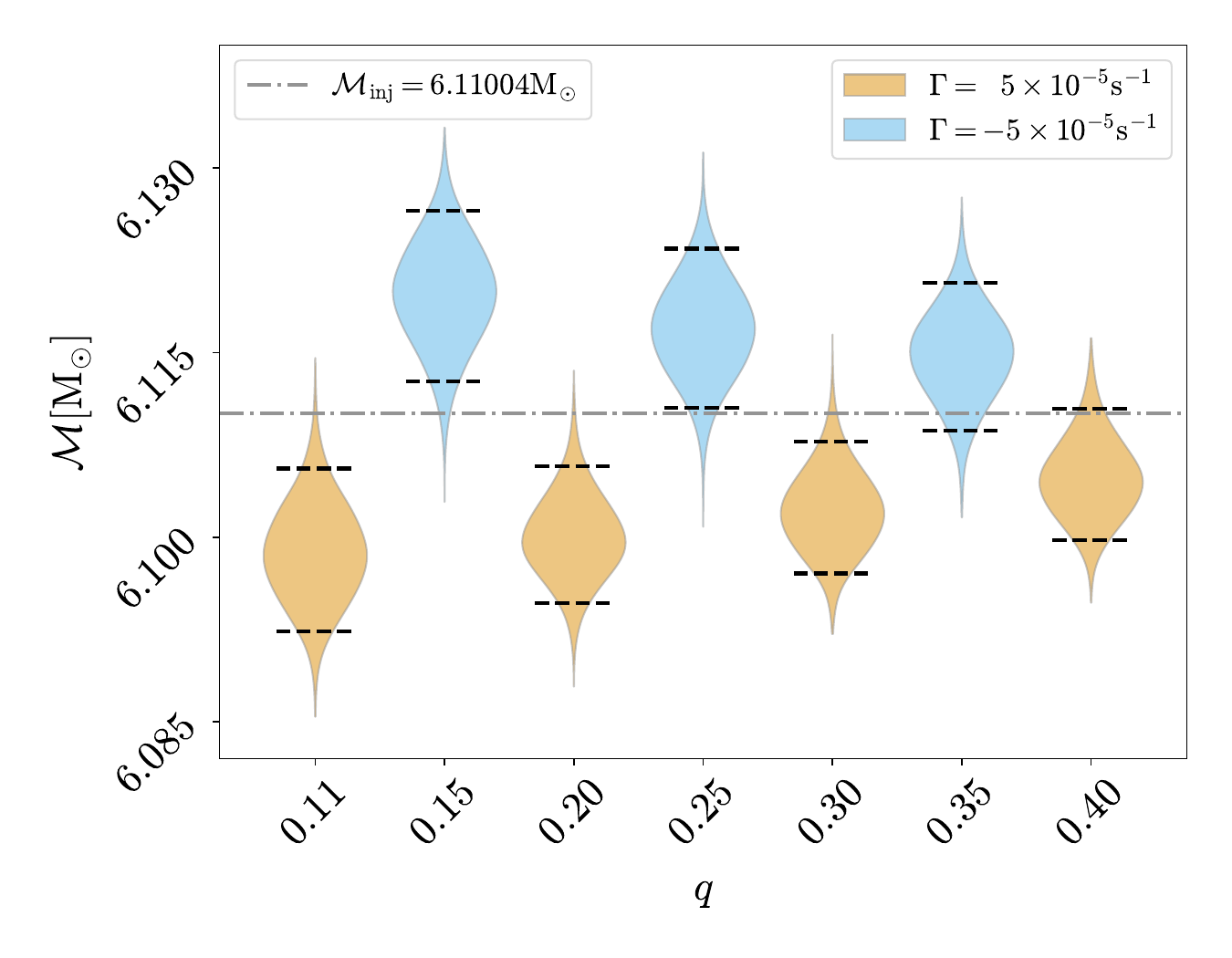}
    \includegraphics[width=0.485\linewidth]{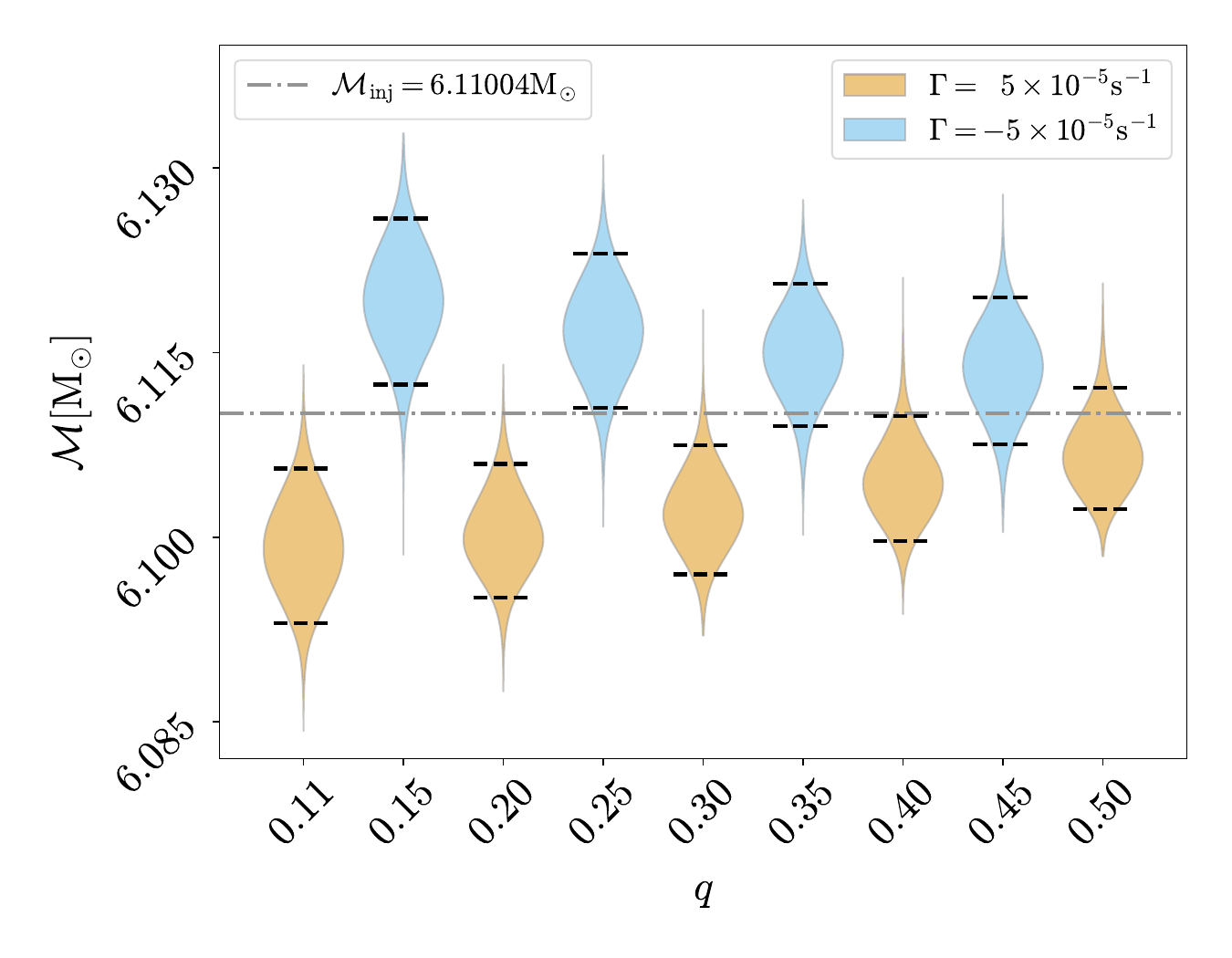}
    \includegraphics[width=0.485\linewidth]{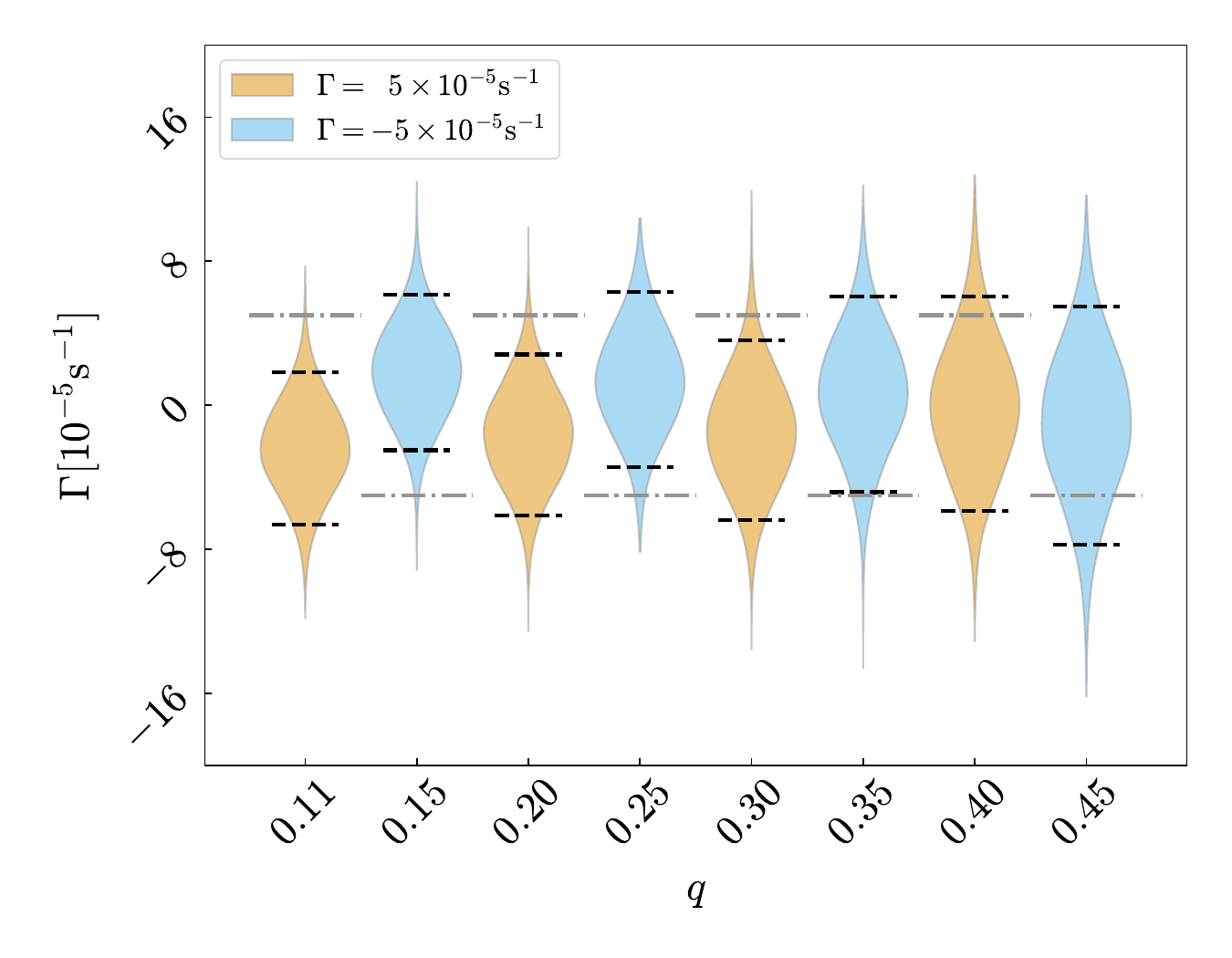}
    \includegraphics[width=0.485\linewidth]{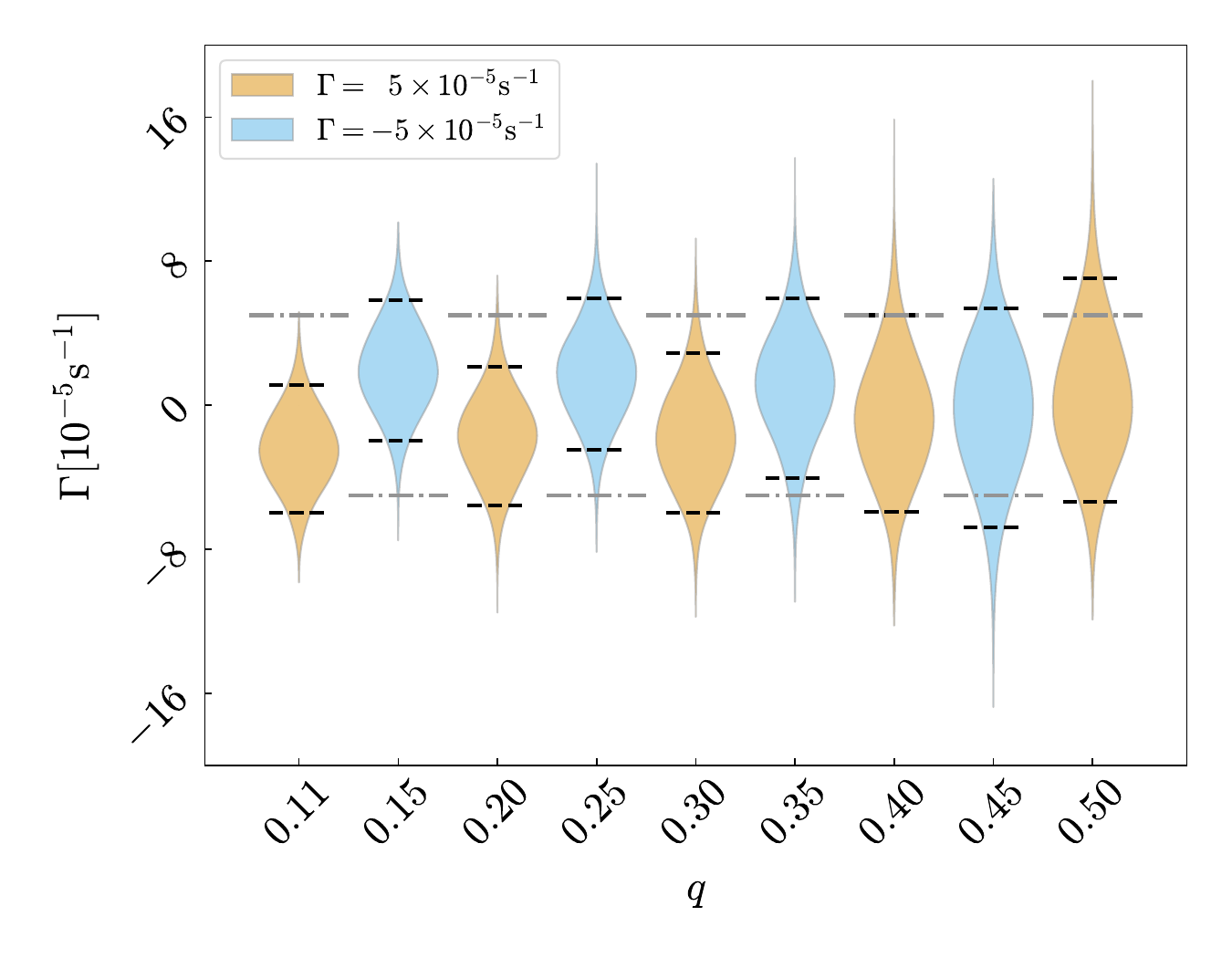}
     
    \caption{The \textit{left panel} represents the posteriors of $\mathcal{M}$ ({\it top panel}) and $\Gamma$ ({\it bottom panel}) when the recovery waveform has (2,2) mode LOSA correction to all modes, while the \textit{right panel} shows the posteriors of the same when the recovery waveform has (2,2) mode LOSA correction only to the (2,2) mode, while the subdominant modes remain unaffected. The dashed lines represent the 90\% CI, while the dash-dot lines represent the injected values.}
    \label{fig: impact_mr_small_losa}
\end{figure*}

\subsection{Impact of Mass Ratio on Biases}\label{mass_ratio_impact}
Having shown the biases in Sections~\ref{sec: all_modes_bias} and~\ref{sec: only_22_mode_bias} due to incorrect application of LOSA phase corrections onto the PE for a GW190814-like signal, in this section, we study how changing the mass ratio affects the PE. This is because, for high mass ratios, the bias is expected to disappear because of a significant reduction in the contribution of the subdominant modes to the total SNR. To demonstrate this, we vary the mass ratio of a GW190814-like signal with all other parameters remaining the same. The injection and recovery settings are the same as those used in Sections~\ref{sec: all_modes_bias} and~\ref{sec: only_22_mode_bias}. Here we change the priors on $q$ to $\mathcal{U}(0.01, 1.0)$. 

Figure~\ref{fig: impact_mr_small_losa} shows the posteriors of $\mathcal{M}$ and $\Gamma$ for different mass ratios. The \textit{left} panel in both plots corresponds to the bias resulting from the implementation of the corrections following Section~\ref{sec: all_modes_bias} and the \textit{right} panel corresponds to the same bias resulting from the implementation of the corrections following Section~\ref{sec: only_22_mode_bias}. The blue posteriors correspond to $\Gamma = -5 \times 10^{-5} {\rm s^{-1}}$, while the orange ones correspond to $\Gamma = 5 \times 10^{-5} {\rm s^{-1}}$. We find that $q$ is recovered within the 90\% CI for most of the cases except for $q = 0.11$, where it is also very close to the 90\% CI edge (see Figure~\ref{fig: impact_mr_small_losa_q} of Appendix~\ref{app: add_plots}), while $\Gamma$ posteriors are consistent with zero for all cases and hence remain unmeasurable across mass ratios, contrary to what we see with the correct implementation, where it was measured for GW190814-like signal (see Figures~\ref{fig: self_consist} and~\ref{fig: self_consist_v2}). We further find that the bias in $\mathcal{M}$ disappears after $q \sim 0.25$ for negative LOSA injection and after $q \sim 0.4$ for positive LOSA injection when recovery settings are similar to Sections~\ref{sec: all_modes_bias} and~\ref{sec: only_22_mode_bias}.

\begin{figure*}[ht!]
    \centering
    \includegraphics[width=0.485\linewidth]{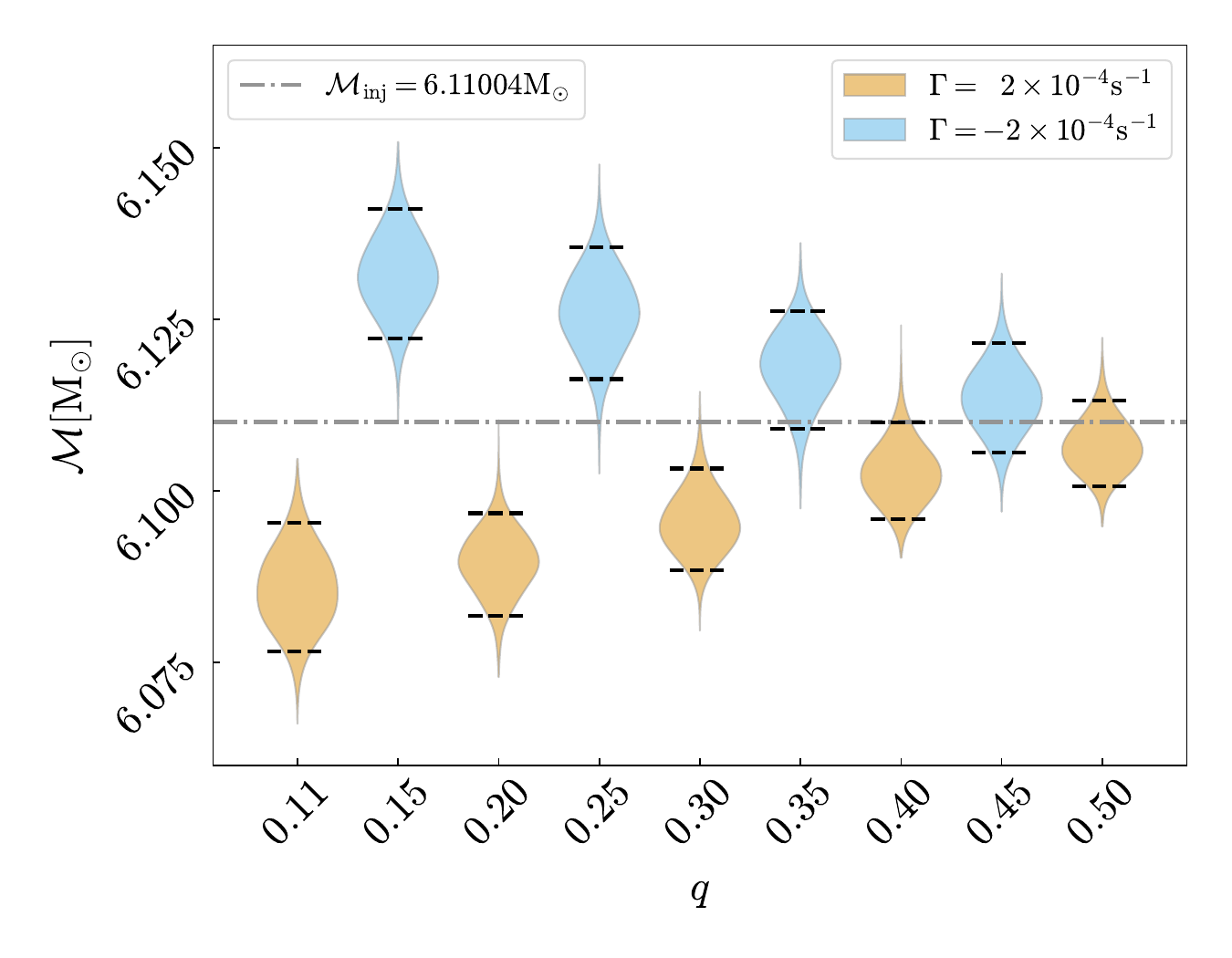}
    \includegraphics[width=0.485\linewidth]{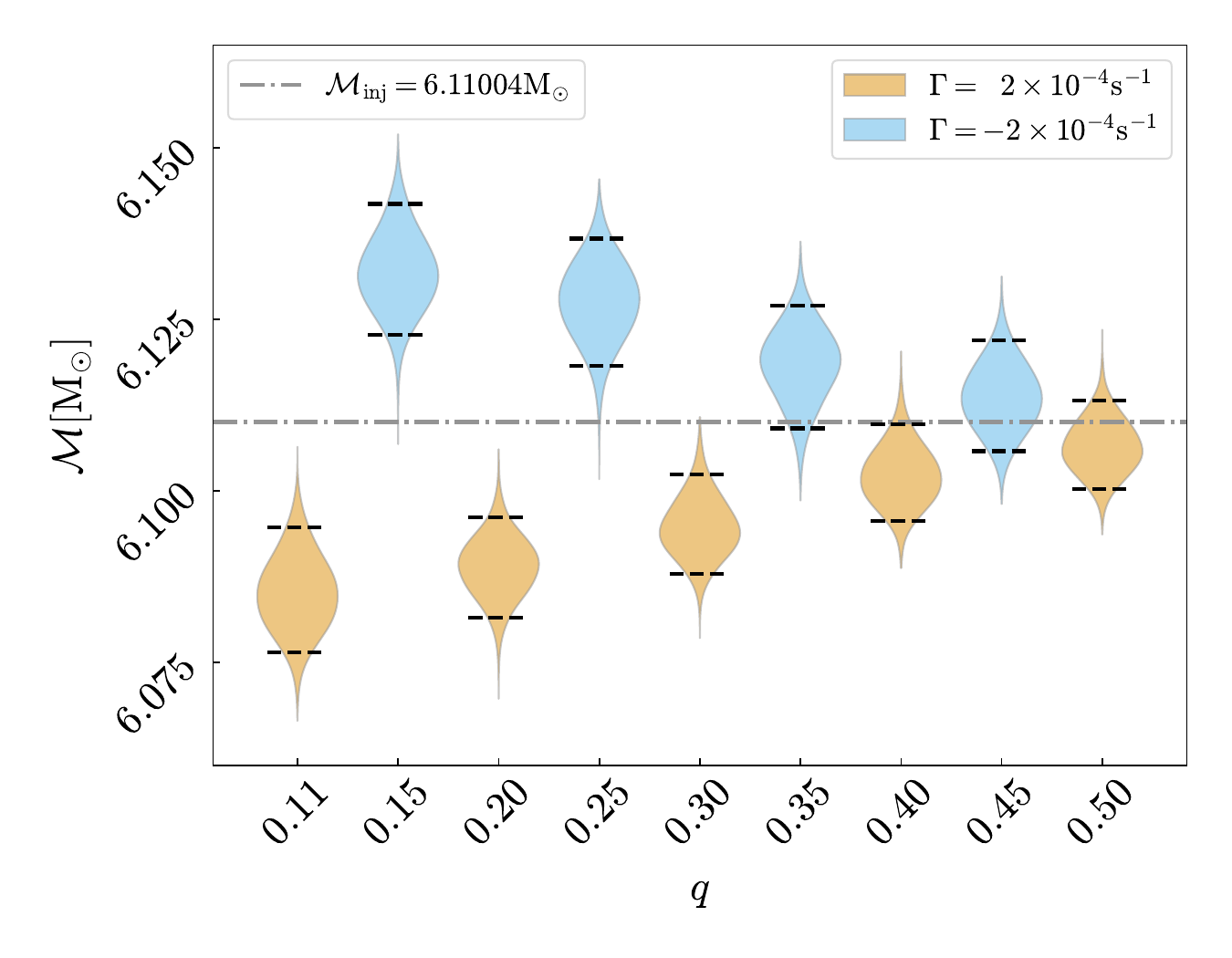}
    \includegraphics[width=0.485\linewidth]{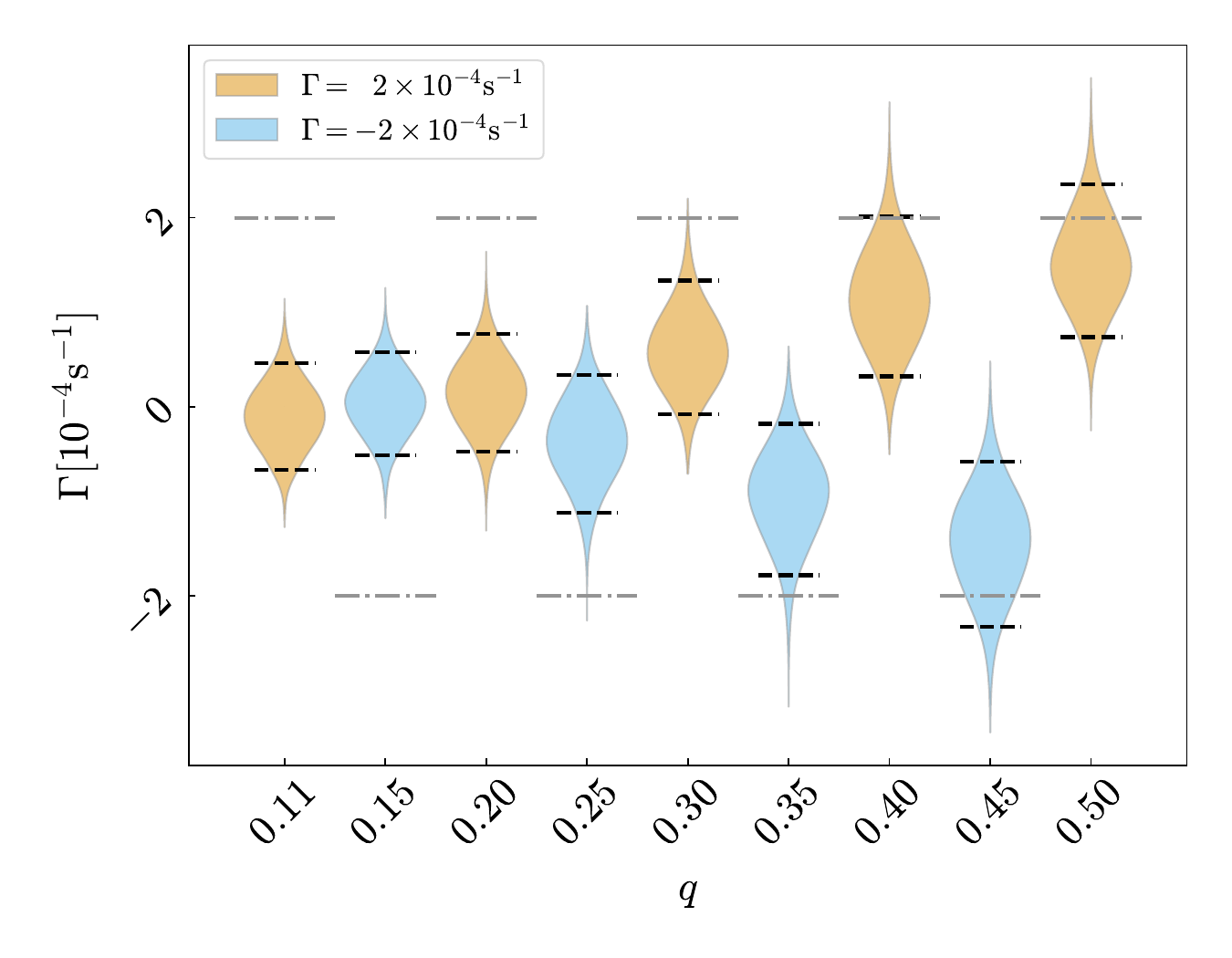}
     \includegraphics[width=0.485\linewidth]{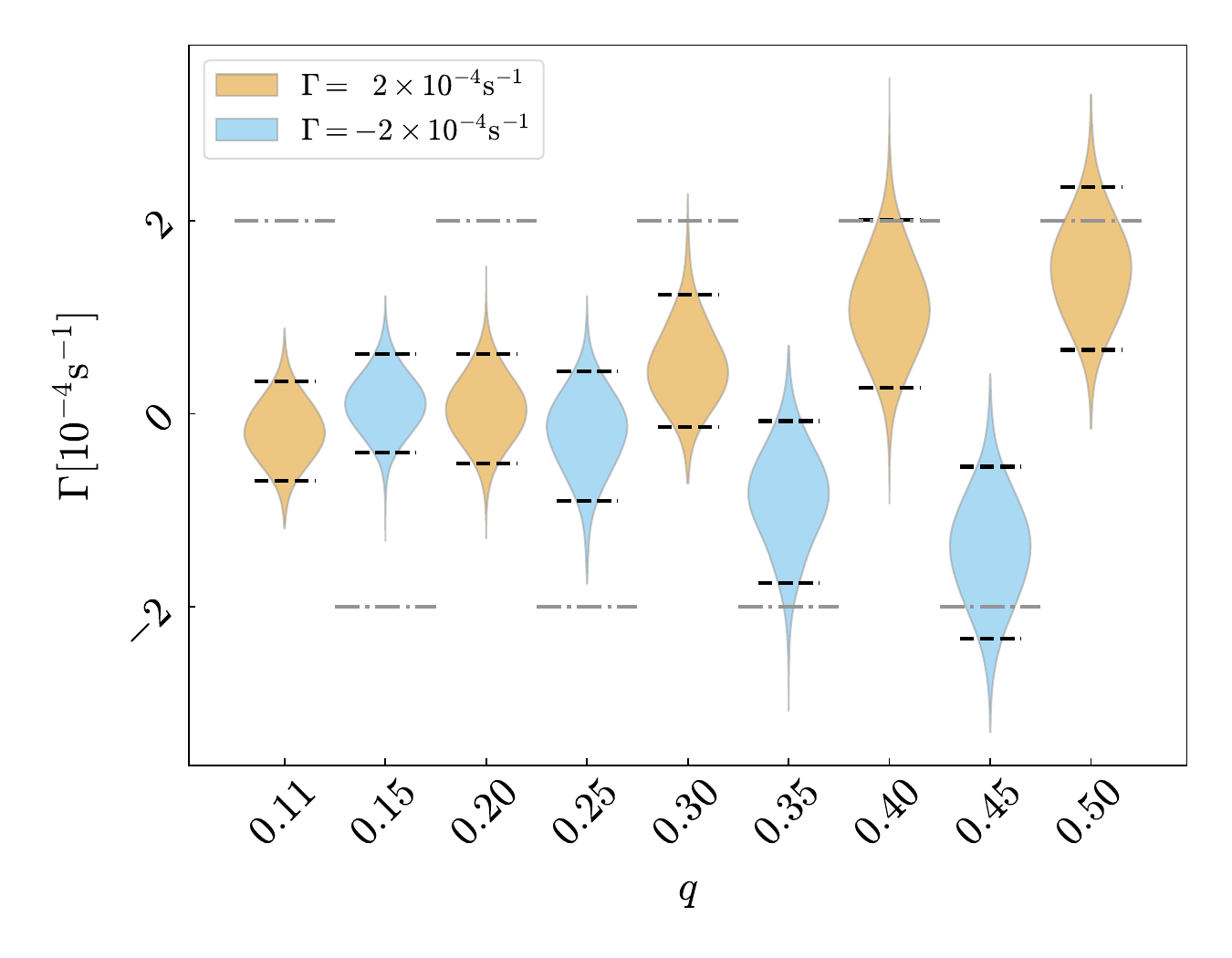}
    \includegraphics[width=0.485\linewidth]{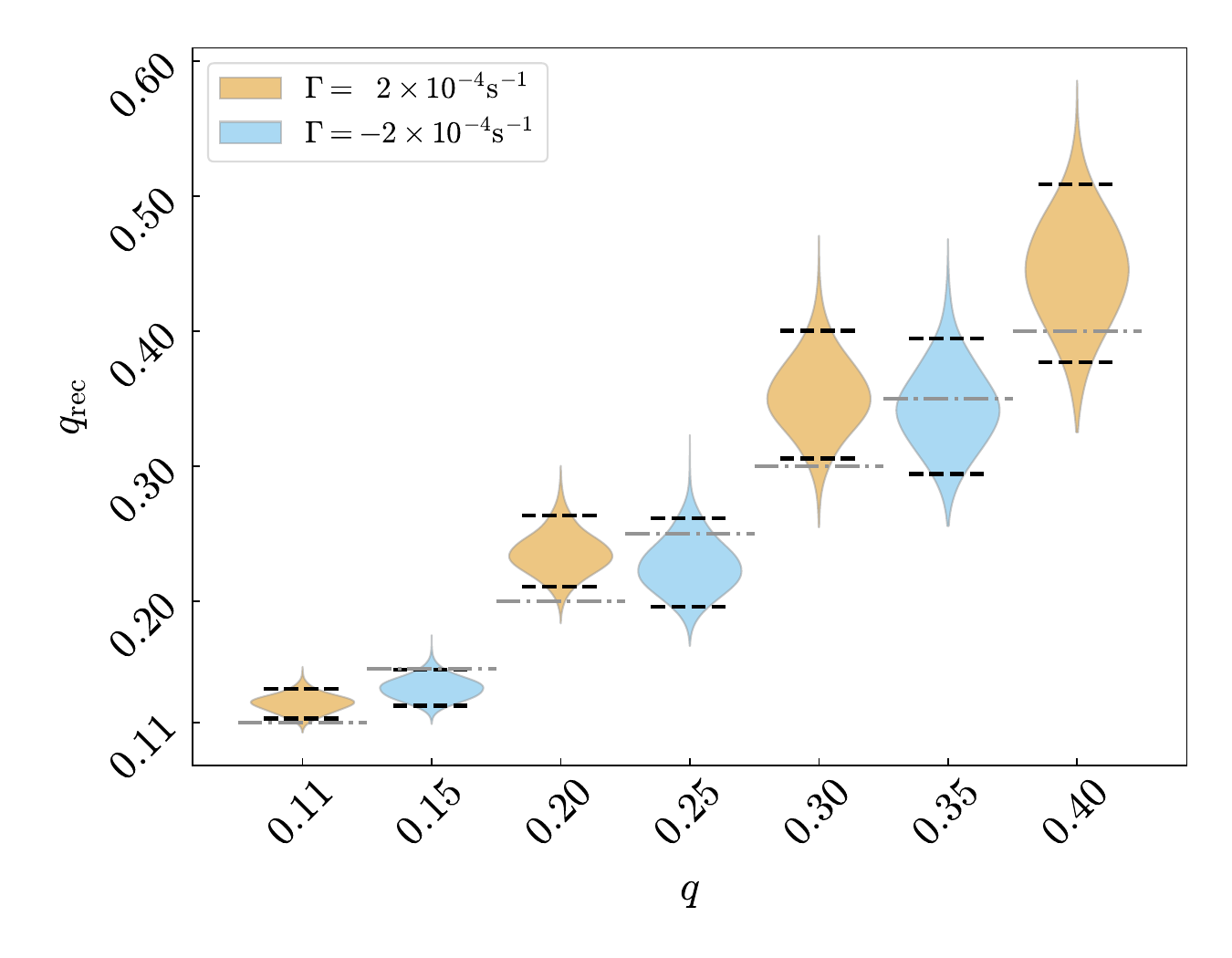}
    \includegraphics[width=0.485\linewidth]{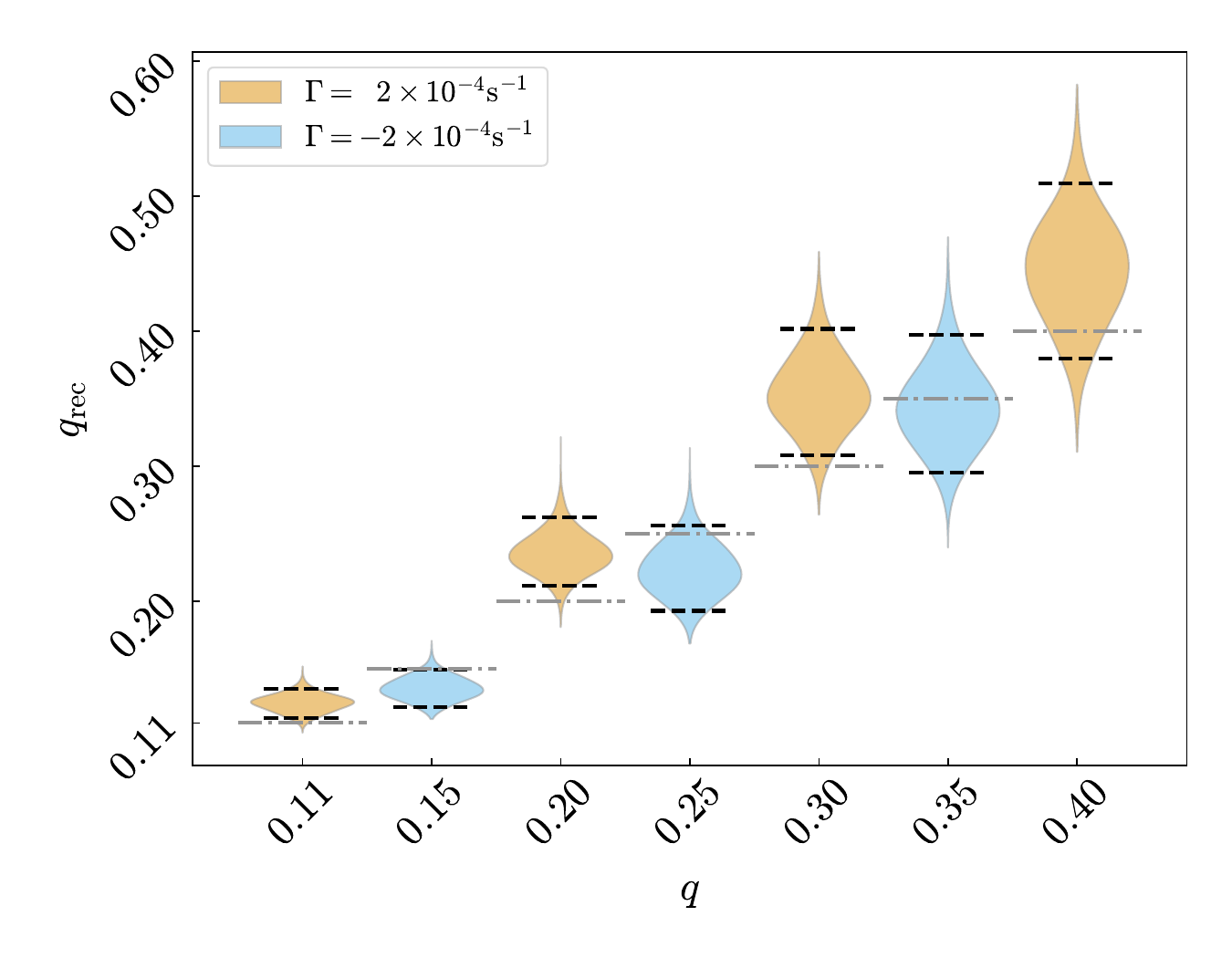}

    \caption{The \textit{left panel} represents the posteriors of $\mathcal{M}$ ({\it top panel}), $\Gamma$ ({\it middle panel}), and $q$ ({\it bottom panel}) when the recovery waveform has (2,2) mode LOSA correction to all subdominant modes, while the \textit{right panel} shows the posteriors of the same when the recovery waveform has (2,2) mode LOSA correction only to the (2,2) mode, while the subdominant modes remain unaffected. The dashed lines represent the 90\% CI, while the dash-dot lines represent the injected values.}
    \label{fig: impact_mr_large_losa}
\end{figure*}

\begin{figure*}[ht!]
    \centering
    \includegraphics[width=0.240\linewidth]{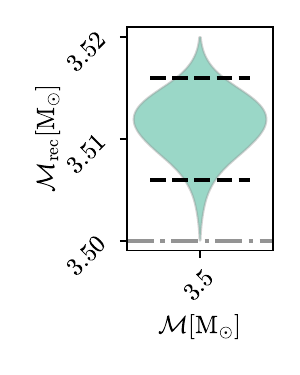}
    \includegraphics[width=0.240\linewidth]{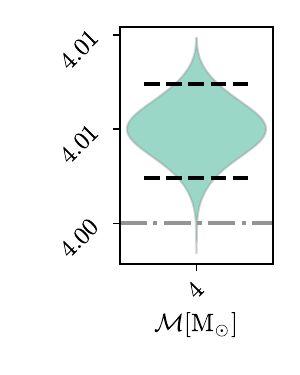}
    \includegraphics[width=0.240\linewidth]{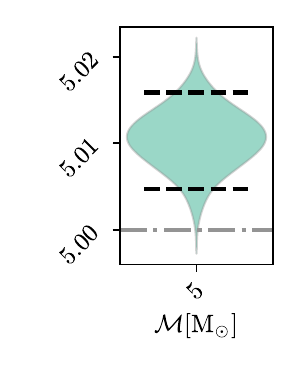}
    \includegraphics[width=0.240\linewidth]{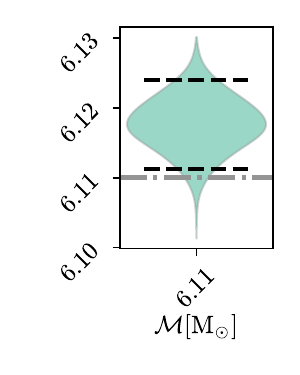}
    
    \includegraphics[width=0.240\linewidth]{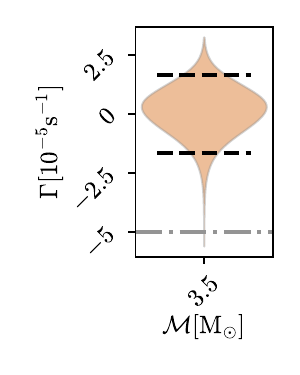}
    \includegraphics[width=0.240\linewidth]{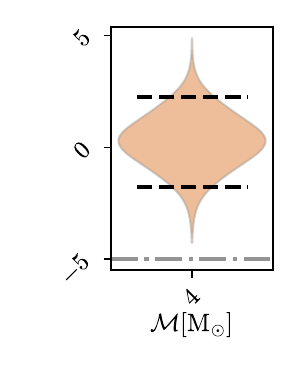}
    \includegraphics[width=0.240\linewidth]{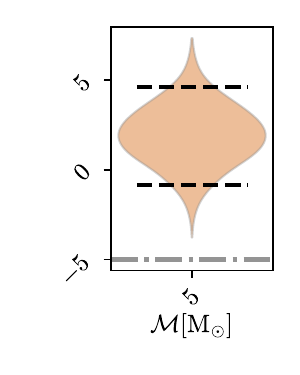}
    \includegraphics[width=0.240\linewidth]{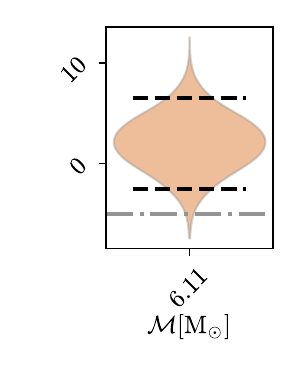}
    
    \includegraphics[width=0.240\linewidth]{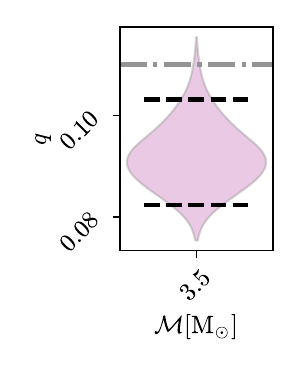}
    \includegraphics[width=0.240\linewidth]{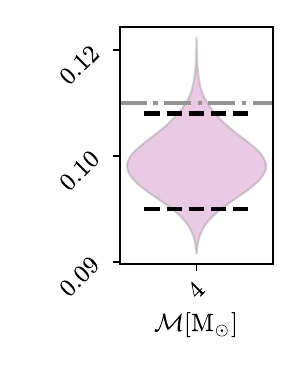}
    \includegraphics[width=0.240\linewidth]{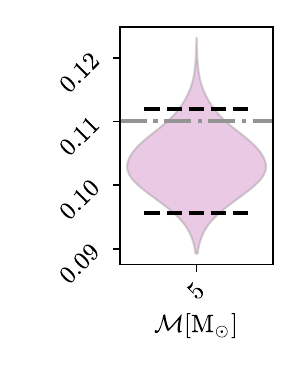}
    \includegraphics[width=0.240\linewidth]{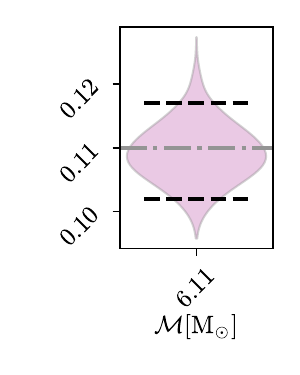}

    \caption{The posteriors of $\mathcal{M}$ ({\it top panel}),  $\Gamma$ ({\it middle panel}), and $q$ ({\it bottom panel}) for different chirp masses and fixed values of $\Gamma = -5 \times 10^{-5} {\rm s^{-1}}$ and $q = 0.11$ when the recovery waveform has the (2,2) mode LOSA correction to all modes. The dashed lines (black) indicate the 90\% CI, while the dash-dot (gray) lines indicate the injected values. }
    \label{fig: changing_Mc_neg_LOSA}
\end{figure*}

 To capture the biases in $q$, we increase the LOSA further and repeat the analysis for $\Gamma = -2 \times 10^{-4} {\rm s^{-1}}$ and $\Gamma = 2 \times 10^{-4} {\rm s^{-1}}$. Figure~\ref{fig: impact_mr_large_losa} shows the posteriors of $\mathcal{M}$, $\Gamma$, and $q$ for this set of runs. The blue posteriors correspond to $\Gamma = - 2 \times 10^{-4} {\rm s^{-1}}$, while the orange ones correspond to $\Gamma = 2 \times 10^{-4} {\rm s^{-1}}$. We find a similar trend in $\Gamma$ posteriors for $q \leq 0.3$ as in Figure~\ref{fig: impact_mr_small_losa}, while it becomes measurable within the 90\% CI for $q \geq 0.4$ but remains biased for $0.3 < q < 0.4$. For $\mathcal{M}$, we find that $\mathcal{M}$ remains biased till $q \sim 0.35$ for negative LOSA injection and till $q \sim 0.4$ for positive LOSA injection.  For $q$, we find that it becomes measurable within the 90\% CI for negative and positive LOSA when $q > 0.15$ and $q > 0.3$, respectively.

Note that to study the sensitivity of the current-generation GW detectors towards these biases, we also repeat this analysis for the O4 detector sensitivities as well. Figure~\ref{fig: bias_O4} in Appendix~\ref{app: bias_O4} shows the results for $\Gamma = \pm 6 \times 10^{-4} \, \rm s^{-1}$ when O4 PSD is used for the analysis. We find that the bias disappears at relatively lower mass ratios even for high $\Gamma$ because of the lower sensitivity of the detectors in comparison to O5. 

\subsection{Impact of Chirp Mass on Biases}\label{chirp_mass_impact}
Having studied the effect of mass ratio on the recovery of the injected parameters with the wrong waveform models, we now move to the effect of varying the chirp mass because for lower chirp mass signals, the bias is expected to become worse because of longer signal duration and a larger number of cycles in band. For this purpose, we again consider a GW190814-like signal, but this time we fix $\Gamma = -5 \times 10^{-5} \, \rm s^{-1}$ and $q = 0.11$, and change the signal duration to $32 \rm s$ while varying the chirp mass. We also change the prior range of the chirp mass so that the upper and lower bounds differ from the injected value by 0.5.

Figure~\ref{fig: changing_Mc_neg_LOSA} shows the corresponding $1d$ posteriors of $\mathcal{M}$, $\Gamma$,and $q$ (see Figure~\ref{fig: changing_Mc_pos_LOSA} of the Appendix~\ref{app: add_plots} for $\Gamma = 5 \times 10^{-5} \, \rm s^{-1}$). We find that $\mathcal{M}$ and $\Gamma$ are biased for all chirp masses considered, while $q$ becomes biased when the chirp mass goes below $5 \, M_{\odot}$. In addition, the bias increases as we decrease the chirp mass, which is expected for two reasons: the first being that LOSA is a $-4$PN order effect and hence becomes stronger as the chirp mass is decreased, and the second being that a smaller chirp mass will have more cycles in the detector bands, resulting in more precise measurements of low-frequency effects like LOSA and, in turn, making the biases more apparent.

\section{Summary and Outlook}\label{sec:conclusions}
In this work, we presented phase and amplitude corrections due to LOSA for higher GW waveform modes. Specifically, we developed an approach based on the conservation of the quantity ${dE}/{f}$ to derive the amplitude corrections, which is different from the one followed in~\cite{Roy:2026mco}; however, our results are in full agreement. Nevertheless, we discuss some conceptual inconsistencies pertaining to the Doppler shift of the GW frequency in  ~\cite{Roy:2026mco}. We explicitly demonstrate that these inconsistencies do not affect the quasi-circular LOSA corrections, but result in non-trivial discrepancies in the eccentric LOSA corrections ~\cite{Ansari:Ecc_2026}. 

We outlined the prescription to apply these corrections to the waveforms and validated it for non-accelerating GW190814-like aligned-spinning injection. We demonstrated that an incorrect application of the (2,2) mode phase and amplitude corrections on all modes, as performed in~\cite{Yang:2024tje}, leads to biased estimates of the parameters for an accelerating system even for $\vert \Gamma \vert = 5 \times 10^{-5} {\rm s^{-1}}$. This suggests that the estimated LOSA claimed in~\cite{Yang:2024tje} for GW190814 is potentially spurious, as also pointed out in~\cite{Tiwari_pipe_2026, Hendriks:2026kys}. We also performed PE on GW190814 and found that the non-accelerating quasi-circular model is preferred over the accelerating one with $\ln (BF) = 1.04$ (see Figure~\ref{fig: real_event} of the Appendix~\ref{app: gw_pe}). Moreover, applying the (2,2) mode phase and amplitude corrections to only the (2,2) mode and keeping the subdominant modes unaffected by LOSA also results in biased parameter estimation. However, we found that, for a GW190814-like system,  neglecting the amplitude corrections does not result in biases, even for the highest LOSA allowed by the constraint, $|\Gamma t_{\rm sd}| \ll 1$.

We studied the effects of varying chirp mass and mass ratio of the CBCs on the biases due to an incorrect application of the LOSA corrections to the waveforms. We found that the biases become progressively less prominent, and eventually become subdominant compared to statistical errors, as the binaries become more symmetric and massive. However, the chirp mass and mass ratios beyond which the biases become negligible depend upon the value of LOSA and the detector sensitivity, as expected. In addition, it also depends upon the sign of the CBC's acceleration.

LOSA is expected to play a crucial role in identifying formation channels of CBCs, potentially on a single-event basis. CBCs in dense stellar environments such as globular clusters or nuclear star clusters are expected to be sources of detectable LOSA, either due to the gravitational potential of the host environment, or because of the presence of third bodies in their vicinity. However, such environments are also expected to produce asymmetric mass CBCs (e.g., due to hierarchical mergers) ~\cite{Tagawa:2019osr, Stevenson:2022djs, Mandel:2021smh, Gerosa:2021mno, KAGRA:2021duu, Abbott_2020_GW190814, Fishbach:2019bbm}. It is therefore crucial to calculate LOSAs self-consistently for GWs with non-trivial higher mode content, without which biased inferences of both LOSA and mass ratio could ensue, corrupting our ability to accurately constrain the CBC's host environment. 

We expect that the LOSA-induced phase and amplitude corrections for GWs with higher modes that we presented in this work can be used to search for signatures of LOSA in quasi-circular CBCs, without modeling biases, particularly for next-generation detectors (Einstein Telescope (ET)~\cite{2010CQGra..27s4002P, 2011CQGra..28i4013H}, Cosmic Explorer (CE)~\cite{2019BAAS...51g..35R, 2022ApJ...931...22S}, DECIGO~\cite{2017JPhCS.840a2010S, 2011PhRvD..83d4011Y, 2017PhRvD..95j9901Y}, LISA~\cite{2012CQGra..29l4016A}, and TianQin~\cite{2016CQGra..33c5010L}) which will be more sensitive to low-frequency effects. It should, however, be noted that dense stellar environments are also expected to produce eccentric CBCs. We refer the reader to \cite{Ansari:Ecc_2026} for the corresponding LOSA-induced corrections pertaining to the $22$ mode and the dominant eccentric harmonic. We also plan to calculate LOSA-driven phase and amplitude corrections for CBCs that are both eccentric and asymmetric-in-masses, in future work.

\begin{acknowledgments}
   We thank Abhishek Sharma for the LVK’s internal Publications and Presentations review of this work and for his careful reading and valuable feedback on the draft. S.J.K. acknowledges support from SERB grants SRG/2023/000419 and MTR/2023/000086. We are grateful for computational resources provided by the Leonard E. Parker Center for Gravitation, Cosmology and Astrophysics at the University of Wisconsin-Milwaukee which is supported by National Science Foundation Grants PHY-1626190 and PHY-2110594 and the IUCAA LDG cluster Sarathi. This material is based upon work supported by NSF's LIGO Laboratory which is a major facility fully funded by the National Science Foundation under Grants PHY-0757058 and PHY-0823459. 
\end{acknowledgments}

\textit{Software}: \texttt{NumPy} \cite{vanderWalt:2011bqk}, \texttt{SciPy} \cite{Virtanen:2019joe}, \texttt{astropy} \cite{2013A&A...558A..33A, 2018AJ....156..123A}, \texttt{Matplotlib} \cite{Hunter:2007}, \texttt{jupyter} \cite{jupyter}, \texttt{LALSuite}~\cite{2020ascl.soft12021L}, \texttt{Bilby}~\cite{bilby_paper}.
\bibliography{refer}


\begin{figure}[ht!]
    \centering
    \includegraphics[width=0.975\linewidth]{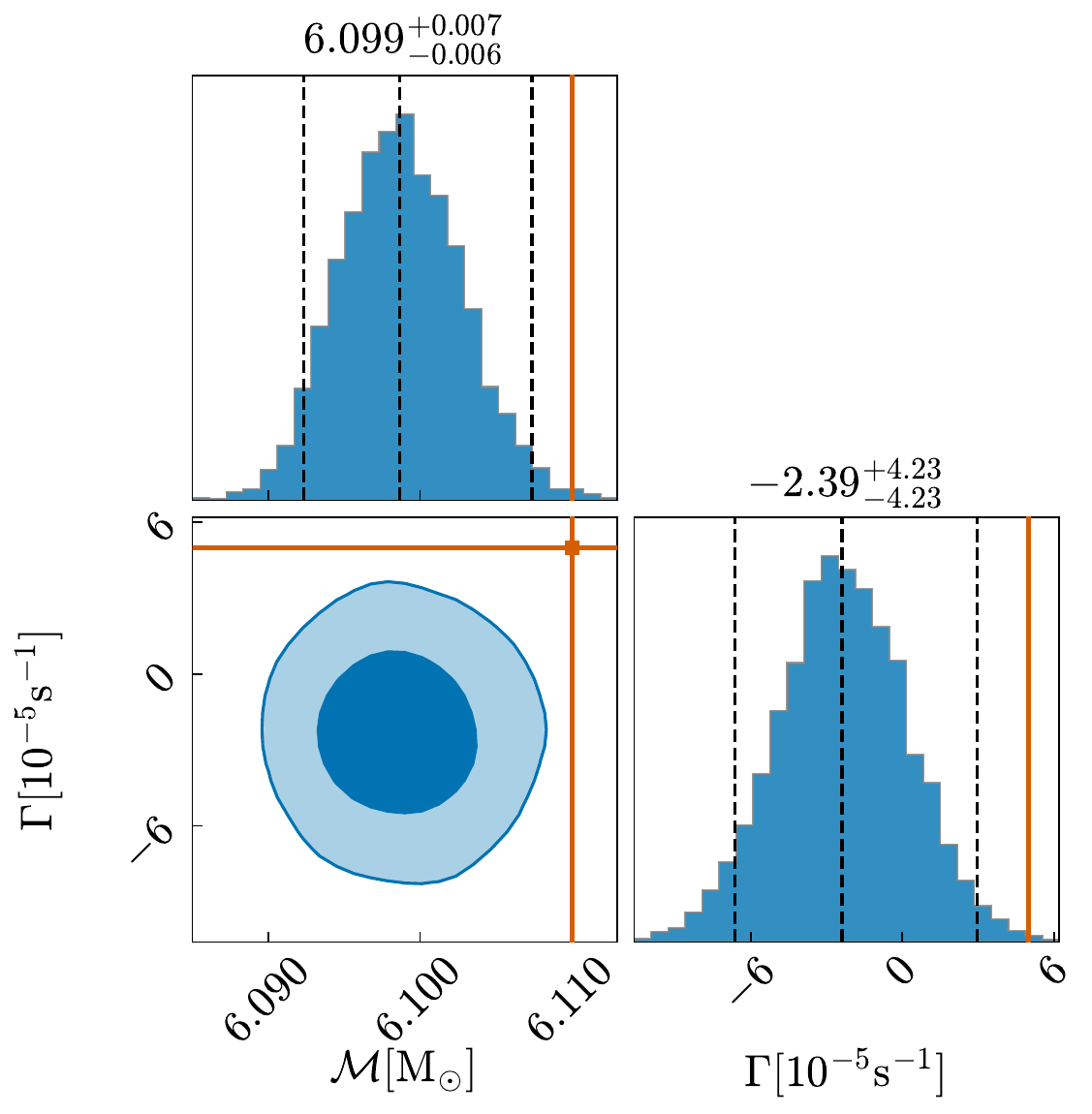}
    \caption{Posteriors of $\mathcal{M}$ and $\Gamma$ for a GW190814-like zero-noise injection, when recovered with a waveform with higher modes and incorrectly applied LOSA phase and amplitude correction of the $(l,m) = (2,2)$ mode to all modes for $\Gamma = 5 \times 10^{-5}{\rm s^{-1}} $ injection.}
    \label{fig: bias_pos_LOSA}

    \centering
    \includegraphics[width=0.975\linewidth]{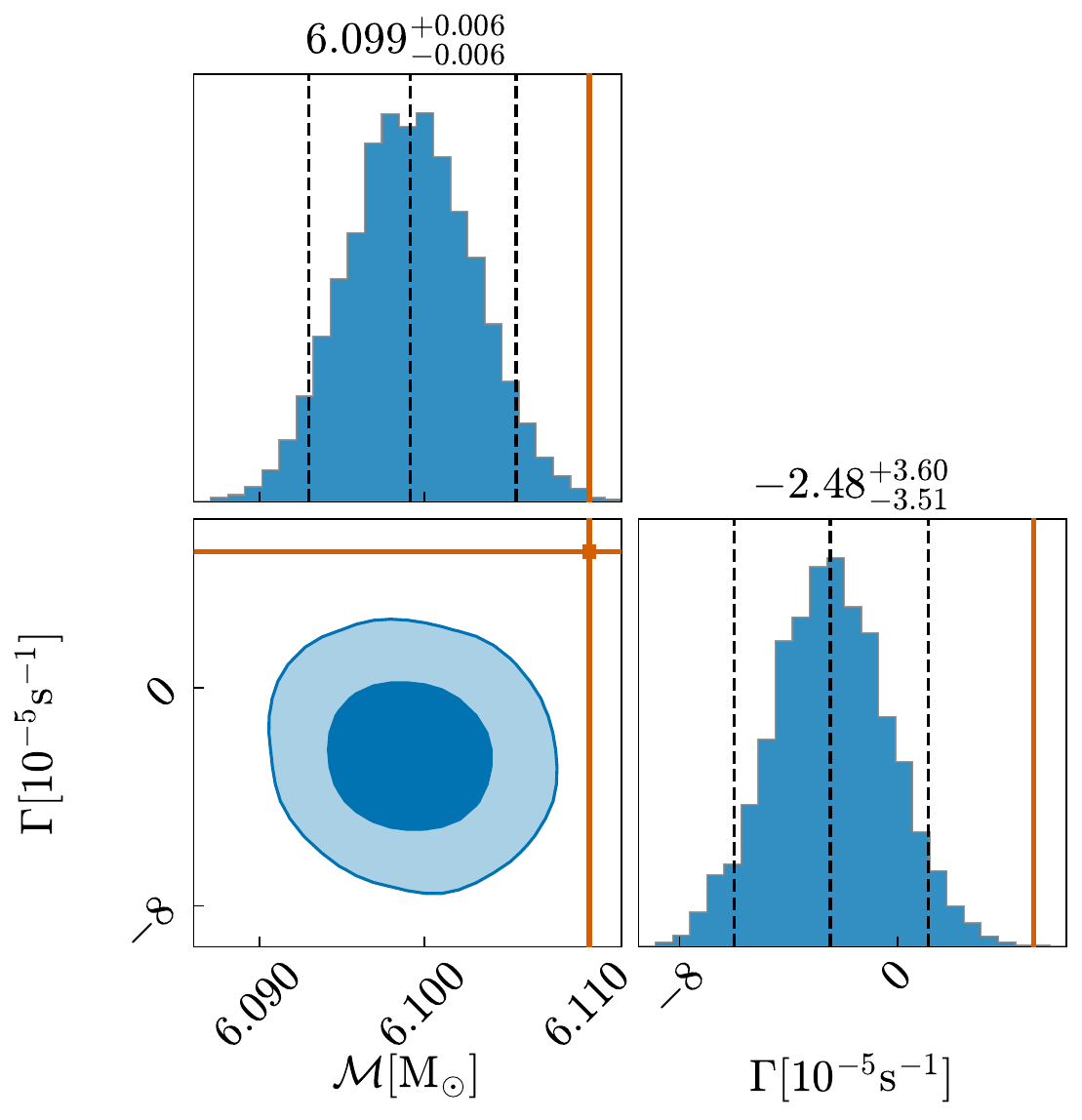}
    \caption{Posteriors of $\mathcal{M}$ and $\Gamma$ for a GW190814-like zero-noise injection when recovered with a waveform with higher modes and incorrectly applied LOSA phase and amplitude correction of the $(l,m) = (2,2)$ mode to only the (2,2) mode, keeping the subdominant modes unaffected, for $\Gamma = 5 \times 10^{-5}{\rm s^{-1}} $} injection.
    \label{fig: bias_pos_LOSA_v2}
\end{figure}

\begin{figure}[ht!]
    \centering
    \includegraphics[width=0.975\linewidth]{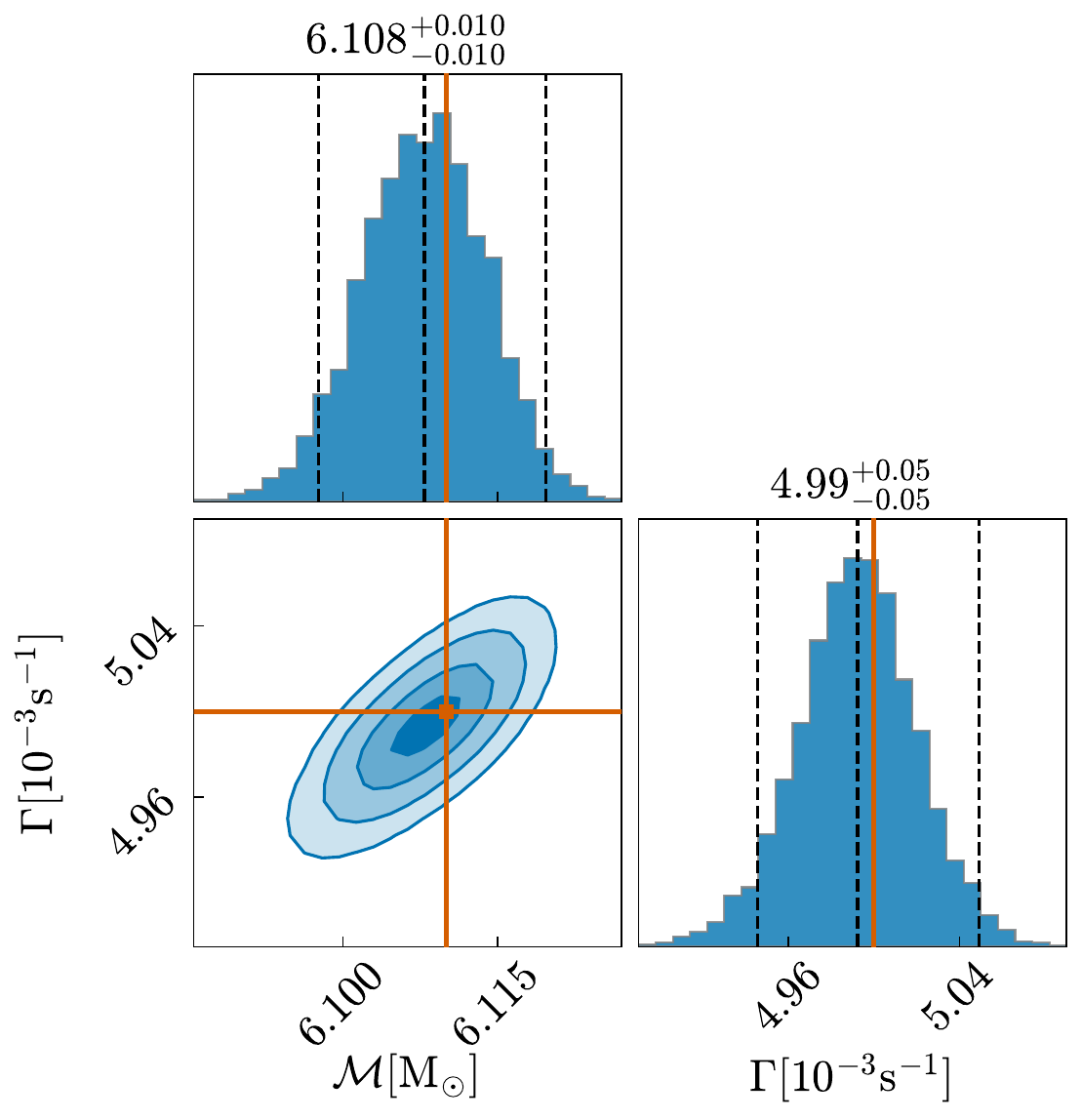}
    \caption{Posteriors of $\mathcal{M}$ and $\Gamma$ for a GW190814-like zero-noise positive LOSA injection when recovered with a waveform without amplitude corrections.}
    \label{fig: amp_corr_v2}

    \centering
    \includegraphics[width=0.975\linewidth]{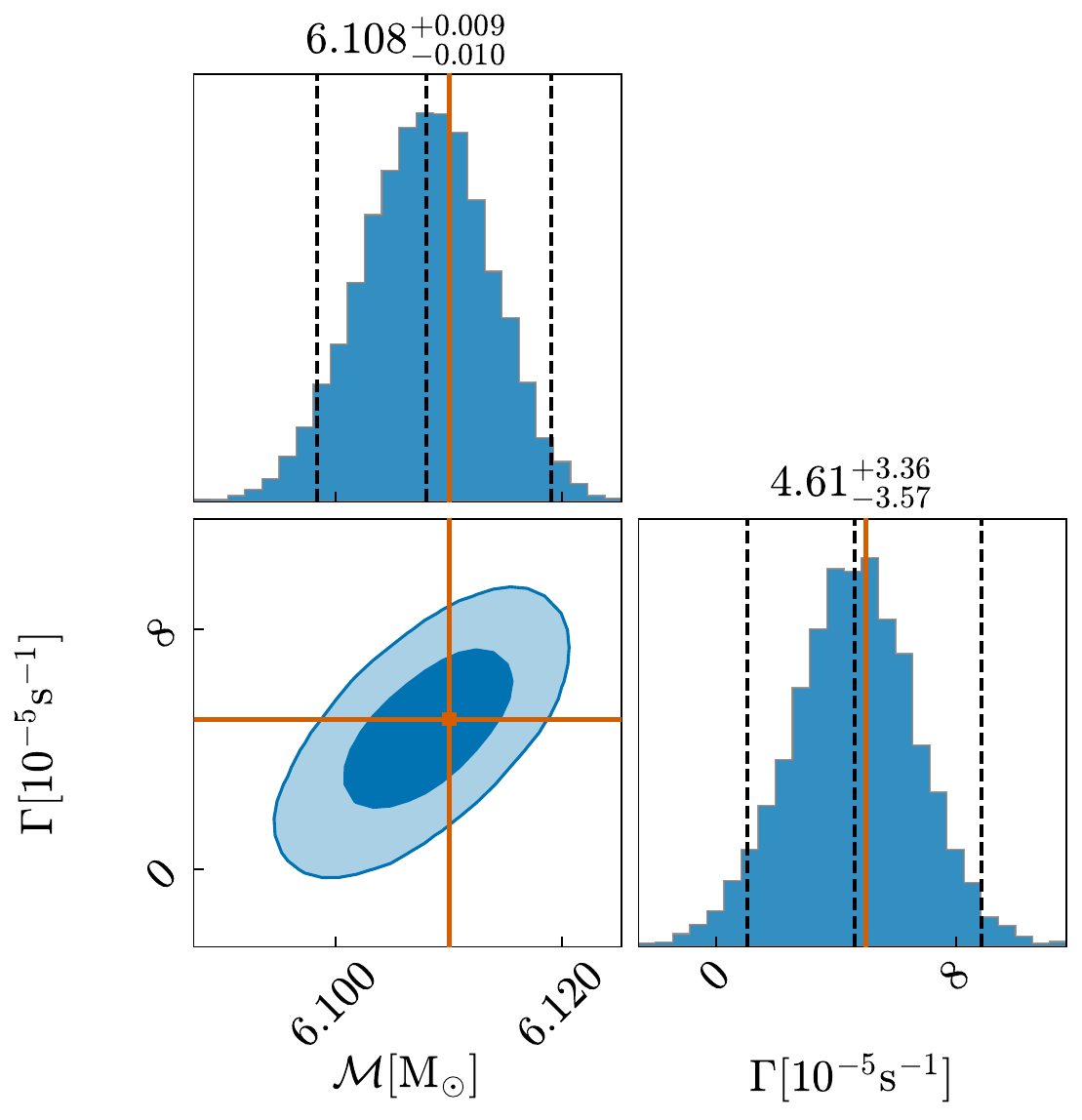}
    \caption{Posteriors of $\mathcal{M}$ and $\Gamma$ for a GW190814-like zero-noise positive LOSA injection, when recovered with a waveform with higher modes and correctly applied LOSA phase and amplitude correction.}
    \label{fig: self_consist_v2}
\end{figure}

\begin{figure*}[ht!]
    \centering
    \includegraphics[width=0.485\linewidth]{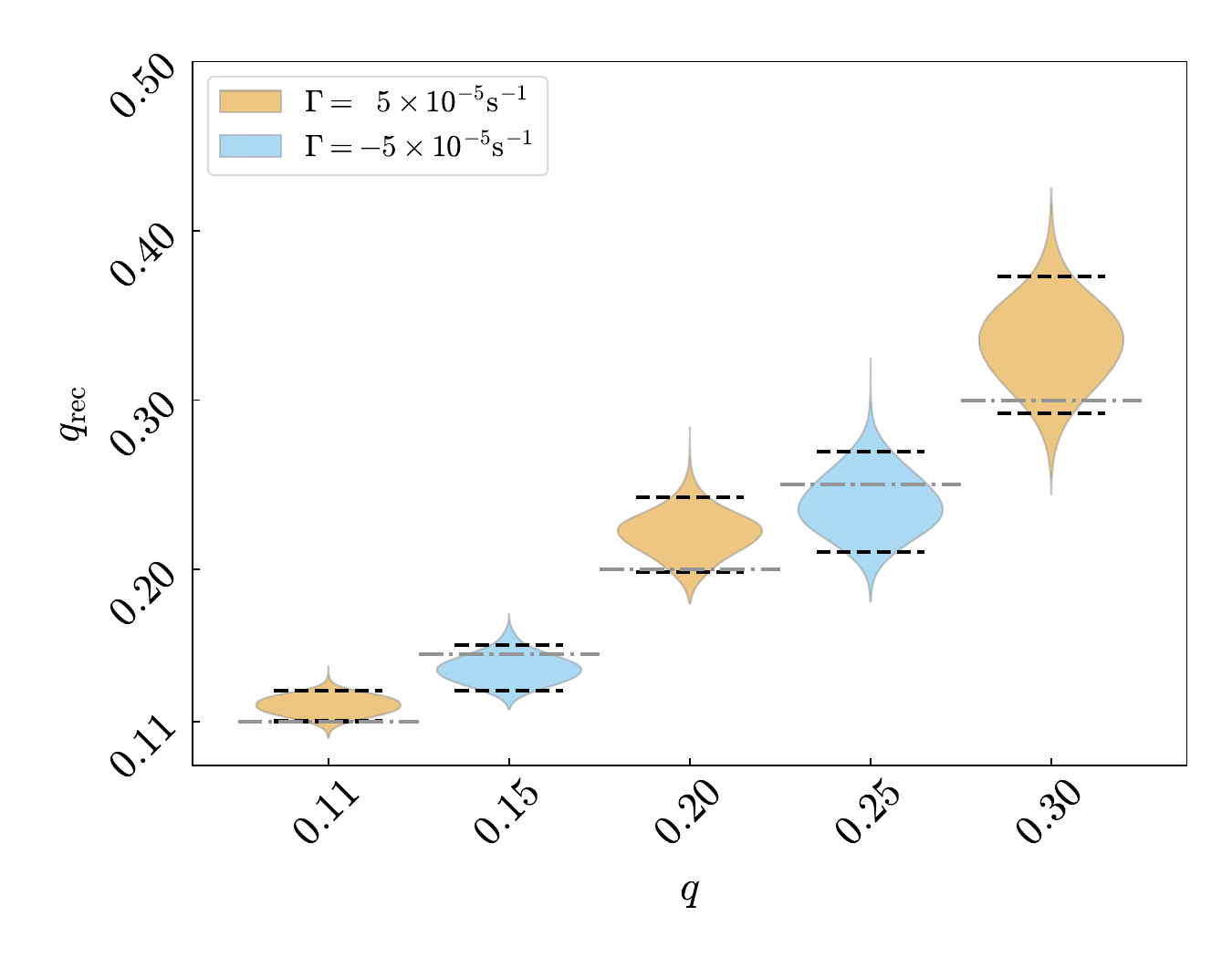}
    \includegraphics[width=0.485\linewidth]{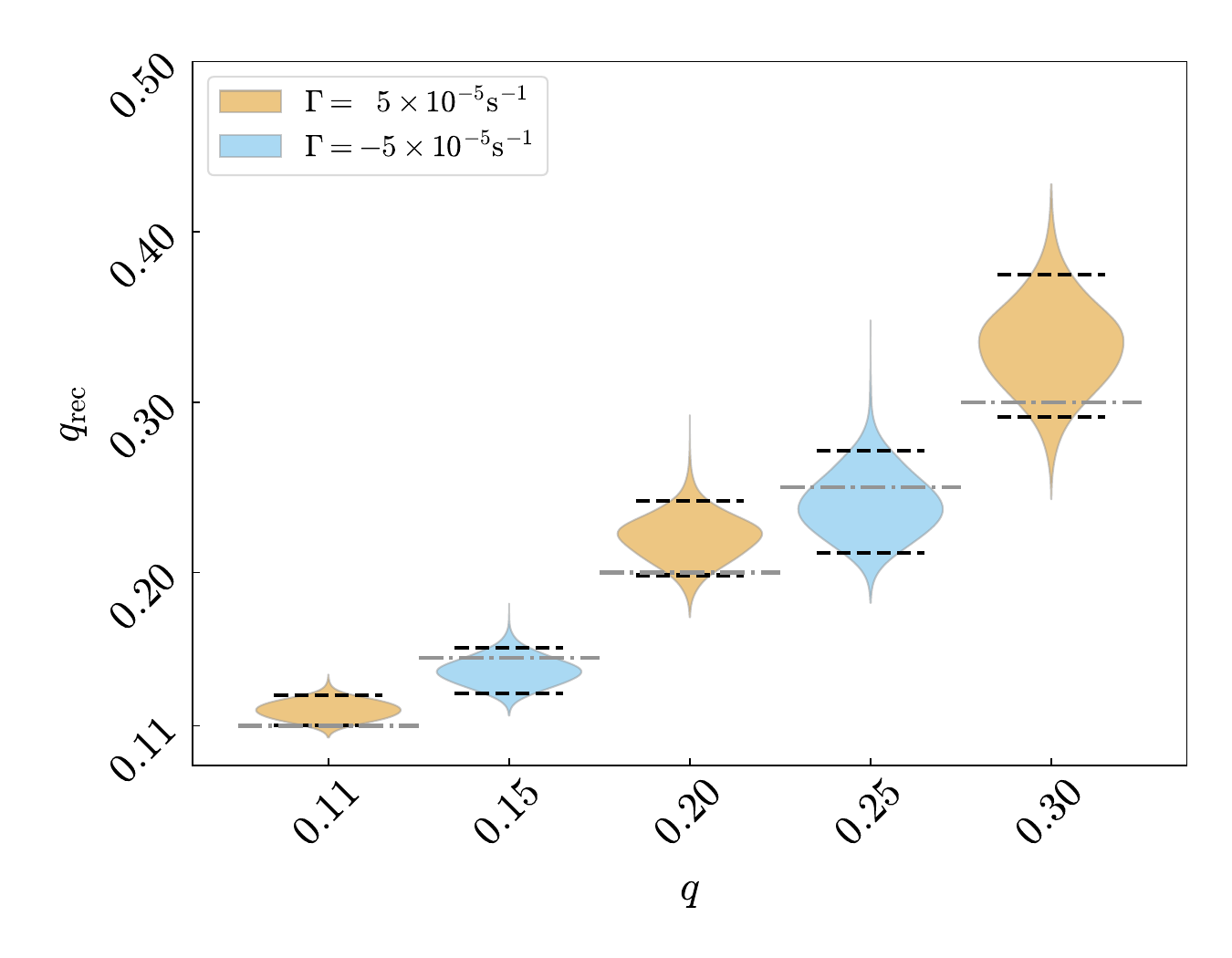}
    \caption{The \textit{left panel} represents the posterior $q$ when the recovery waveform has (2,2) mode LOSA correction to all modes, while the \textit{right panel} shows the posterior of the same when the recovery waveform has (2,2) mode LOSA correction only to the (2,2) mode, while the subdominant modes remain unaffected. }
    \label{fig: impact_mr_small_losa_q}
\end{figure*}

\appendix

\section{Additional Plots}\label{app: add_plots}
In Figures~\ref{fig: bias_pos_LOSA} and~\ref{fig: bias_pos_LOSA_v2}, we show the posteriors of $\mathcal{M}$ and $\Gamma$ for $\Gamma = 5 \times 10^{-5} {\rm s^{-1}}$ corresponding to the biases discussed in Sections~\ref{sec: all_modes_bias} and~\ref{sec: only_22_mode_bias}, respectively. We find that, similar to the negative LOSA injection case, both parameters are biased here as well.

Figure~\ref{fig: amp_corr_v2} shows the posterior of $\mathcal{M}$ and $\Gamma$ for $\Gamma = 5 \times 10^{-3} {\rm s^{-1}}$ when the amplitude correction is neglected in the recovery waveform. We find that ignoring the amplitude corrections does not result in a parameter bias even for a positive LOSA.

Figure~\ref{fig: self_consist_v2} shows the $\mathcal{M}$ and $\Gamma$ posteriors for $\Gamma = 5 \times 10^{-5} {\rm s^{-1}}$ when the phase and amplitude corrections are consistently imposed across the full harmonic structure following the setup described in Section~\ref{sec: self_cons}. We find that the inferred chirp mass and LOSA are consistent with the corresponding injected values within the 90\% CI.

Figure~\ref{fig: impact_mr_small_losa_q} shows the $1d$ violin posteriors on the mass ratio at different injected mass ratio values. The left and right panels correspond to the biases described in Section ~\ref{sec: all_modes_bias} and ~\ref{sec: only_22_mode_bias}. The blue violins correspond to $\Gamma = -5 \times 10^{-5} {\rm s^{-1}}$ and the orange ones to  $\Gamma = 5 \times 10^{-5} {\rm s^{-1}}$. We find that the bias shows up only for the positive LOSA case around $q \sim 0.11$ and disappears for large values of $q$.

\begin{figure*}
    \centering
    \includegraphics[width=0.240\linewidth]{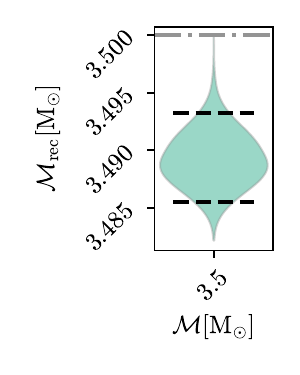}
    \includegraphics[width=0.240\linewidth]{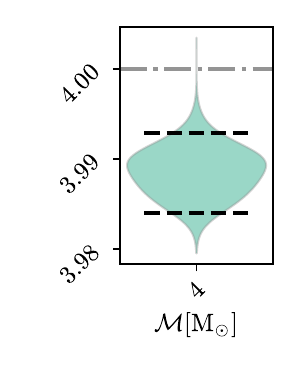}
    \includegraphics[width=0.240\linewidth]{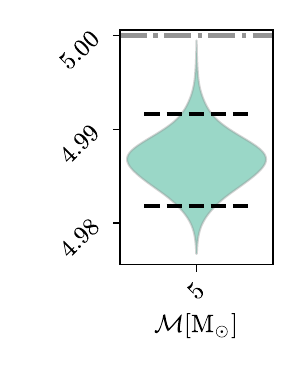}
    \includegraphics[width=0.240\linewidth]{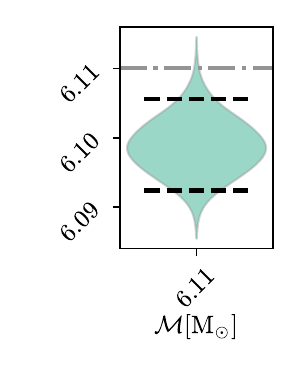}
    
    \includegraphics[width=0.240\linewidth]{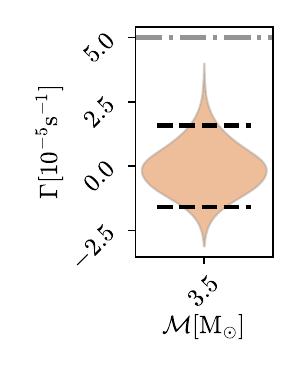}
    \includegraphics[width=0.240\linewidth]{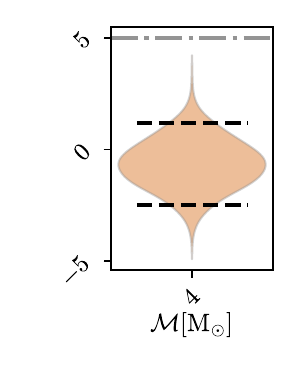}
    \includegraphics[width=0.240\linewidth]{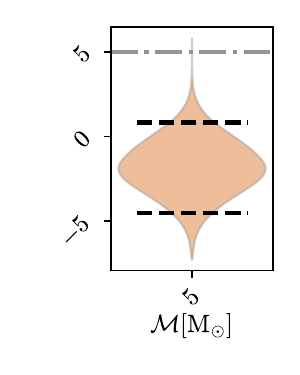}
    \includegraphics[width=0.240\linewidth]{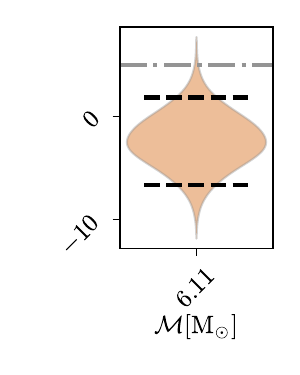}
    
    \includegraphics[width=0.240\linewidth]{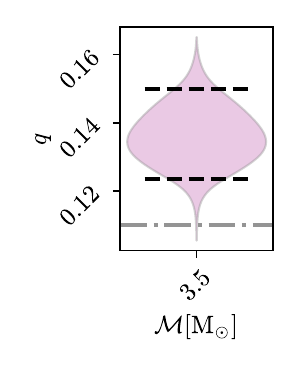}
    \includegraphics[width=0.240\linewidth]{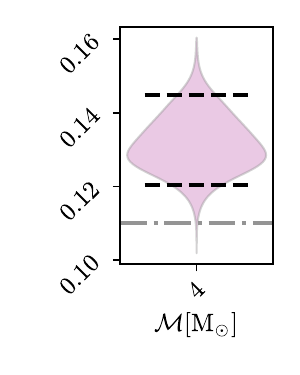}
    \includegraphics[width=0.240\linewidth]{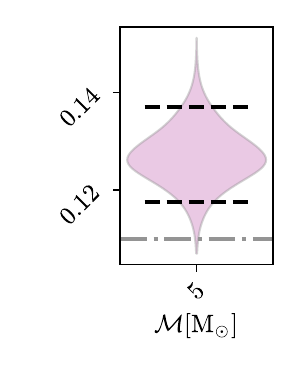}
    \includegraphics[width=0.240\linewidth]{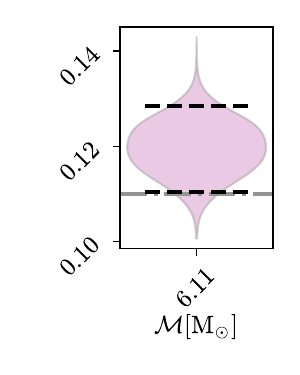}

    \caption{The posteriors of $\mathcal{M}$ ({\it top panel}),  $\Gamma$ ({\it middle panel}), and $q$ ({\it bottom panel}) for different chirp masses and fixed values of $\Gamma = 5 \times 10^{-5} {\rm s^{-1}}$ and $q = 0.11$ when the recovery waveform has the (2,2) mode LOSA correction to all modes. The dashed lines (black) indicate the 90\% CI, while the dash-dot (gray) lines indicate the injected values. }
    \label{fig: changing_Mc_pos_LOSA}
\end{figure*}

Figure~\ref{fig: changing_Mc_pos_LOSA} shows the $1d$ posteriors of $\mathcal{M}$, $\Gamma$,and $q$ for $\Gamma = 5 \times 10^{-5} \, \rm s^{-1}$) for different chirp masses and fixed $\Gamma$ and $q$.

\section{Biases in O4}\label{app: bias_O4}
\begin{figure*}[ht!]
    \centering
    \includegraphics[width=0.485\linewidth]{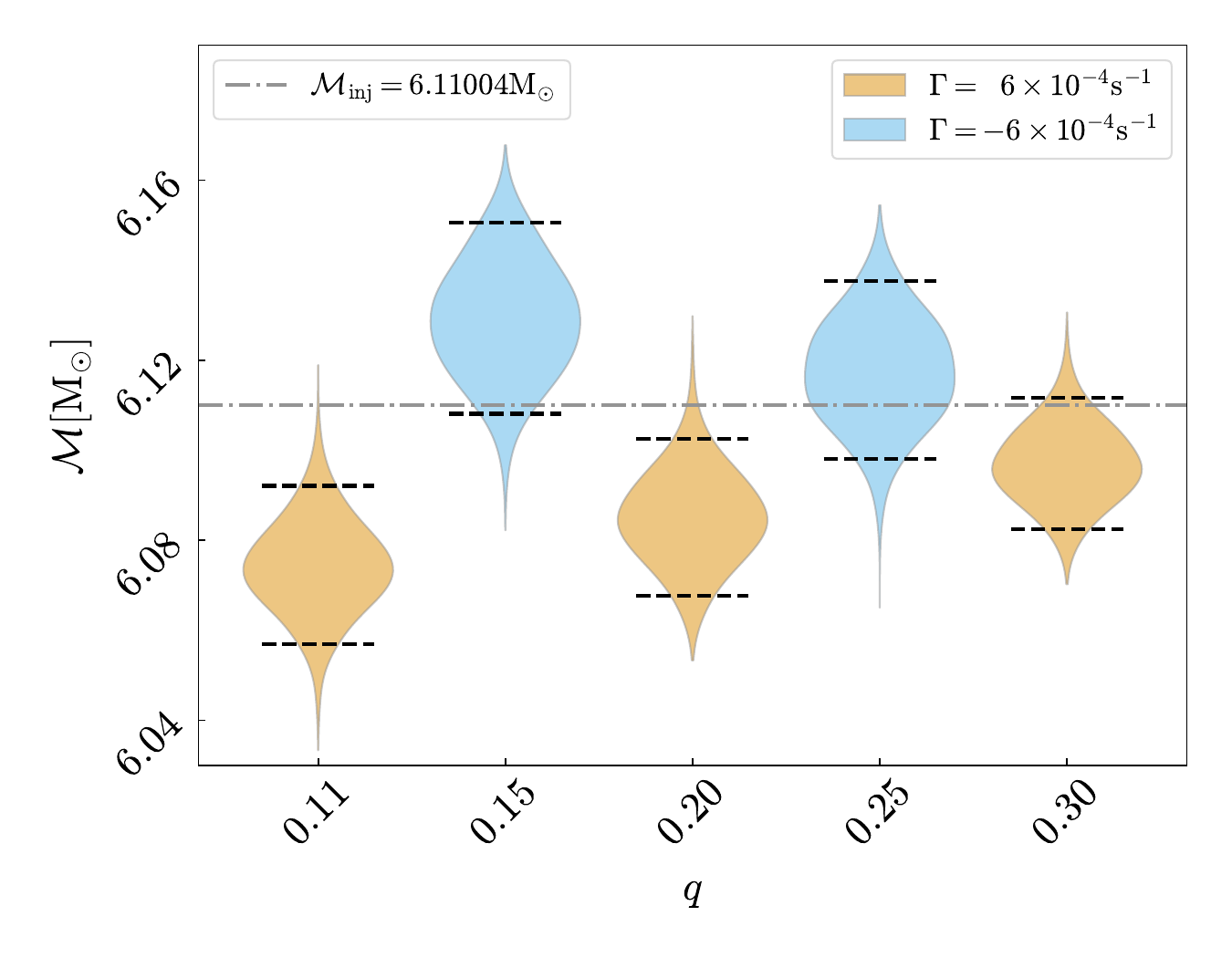}
    \includegraphics[width=0.485\linewidth]{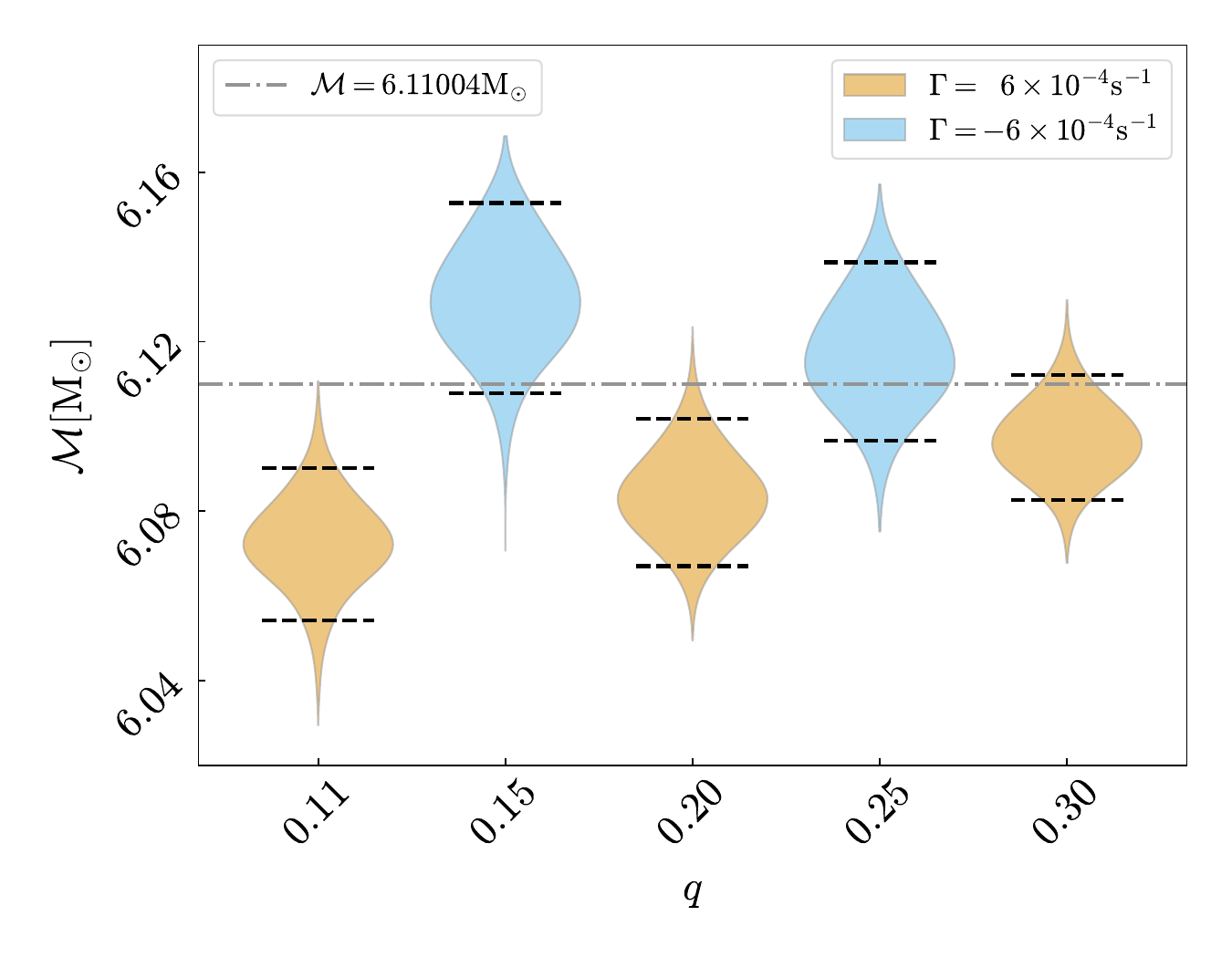}
    \includegraphics[width=0.485\linewidth]{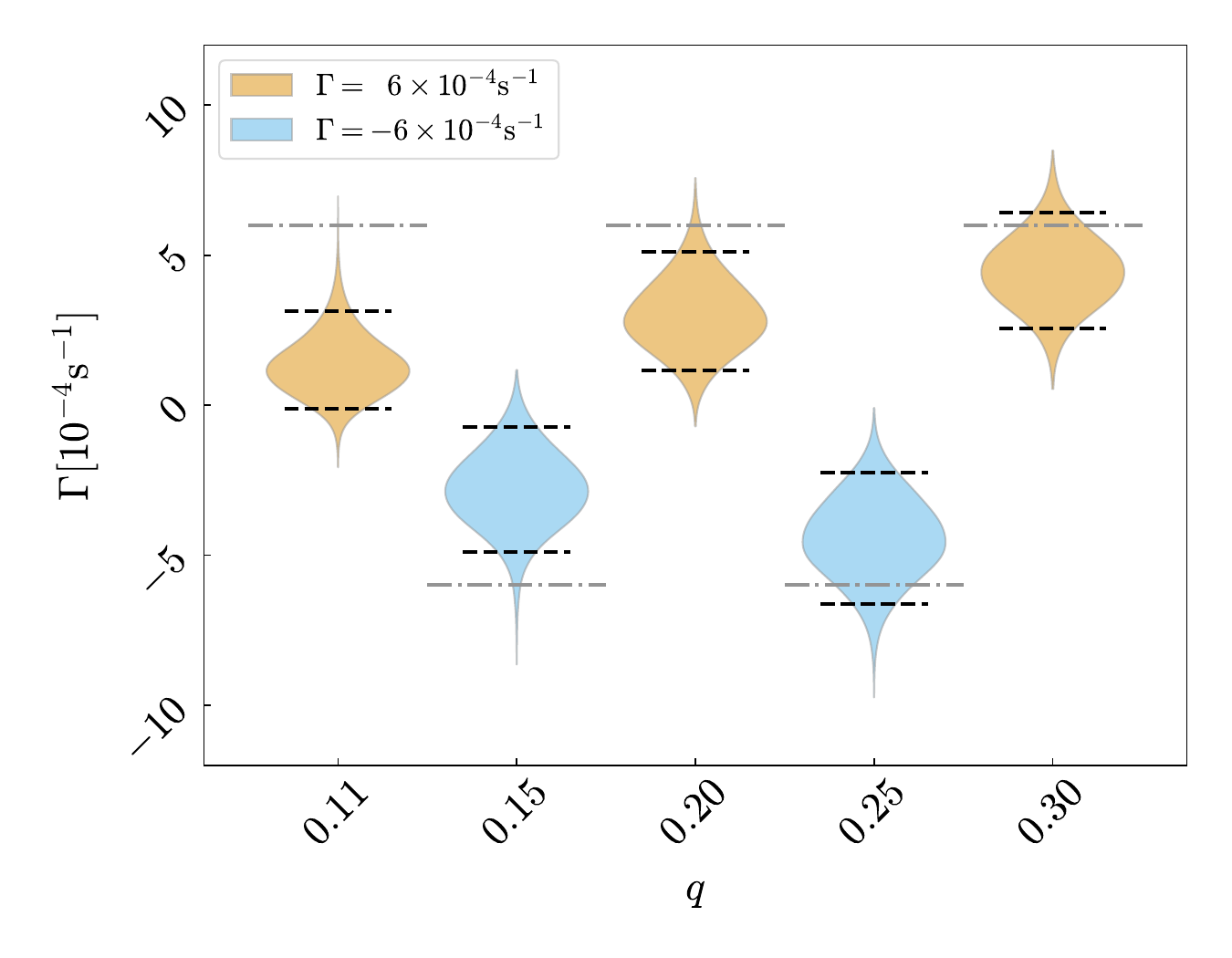}
     \includegraphics[width=0.485\linewidth]{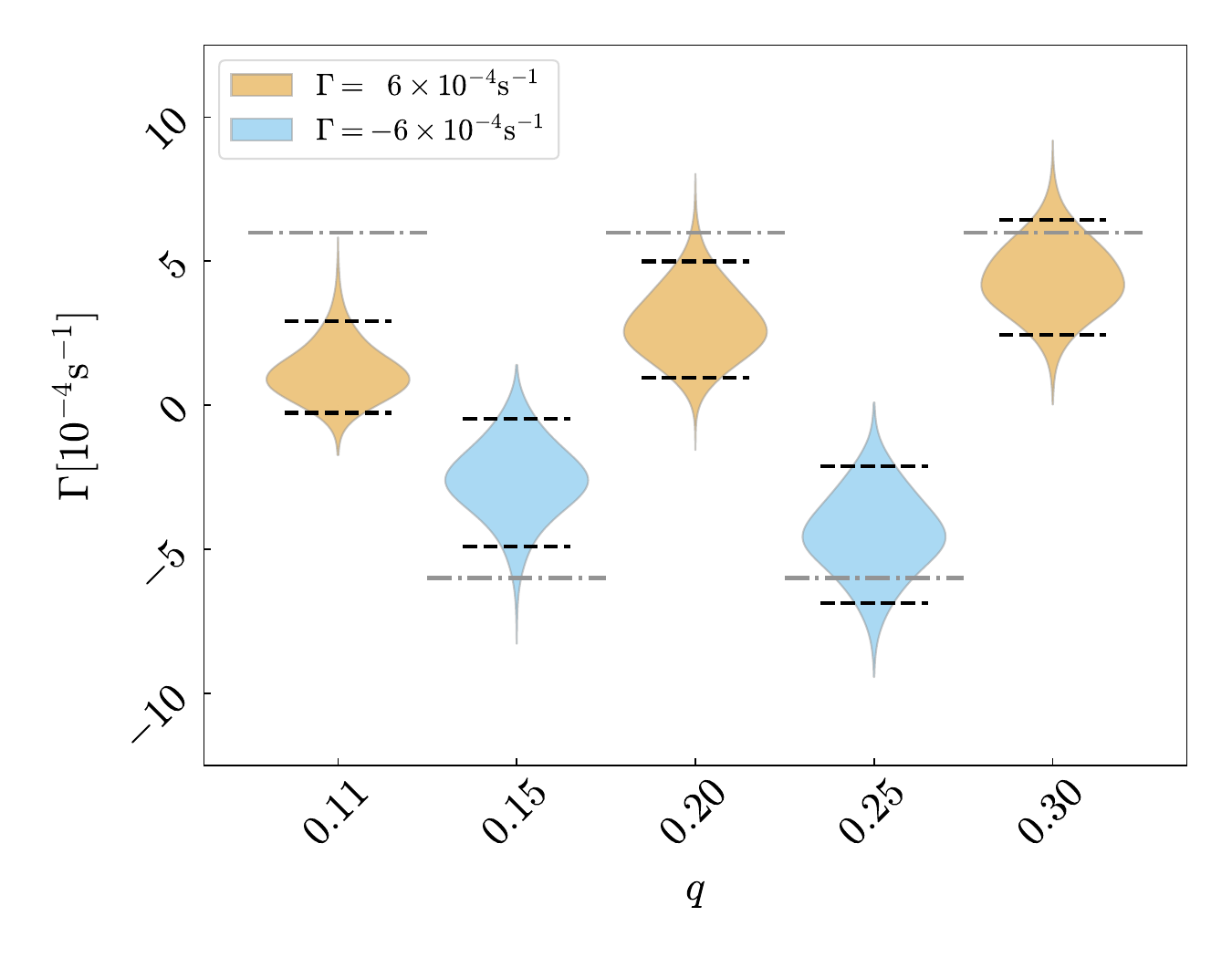}
     \includegraphics[width=0.485\linewidth]{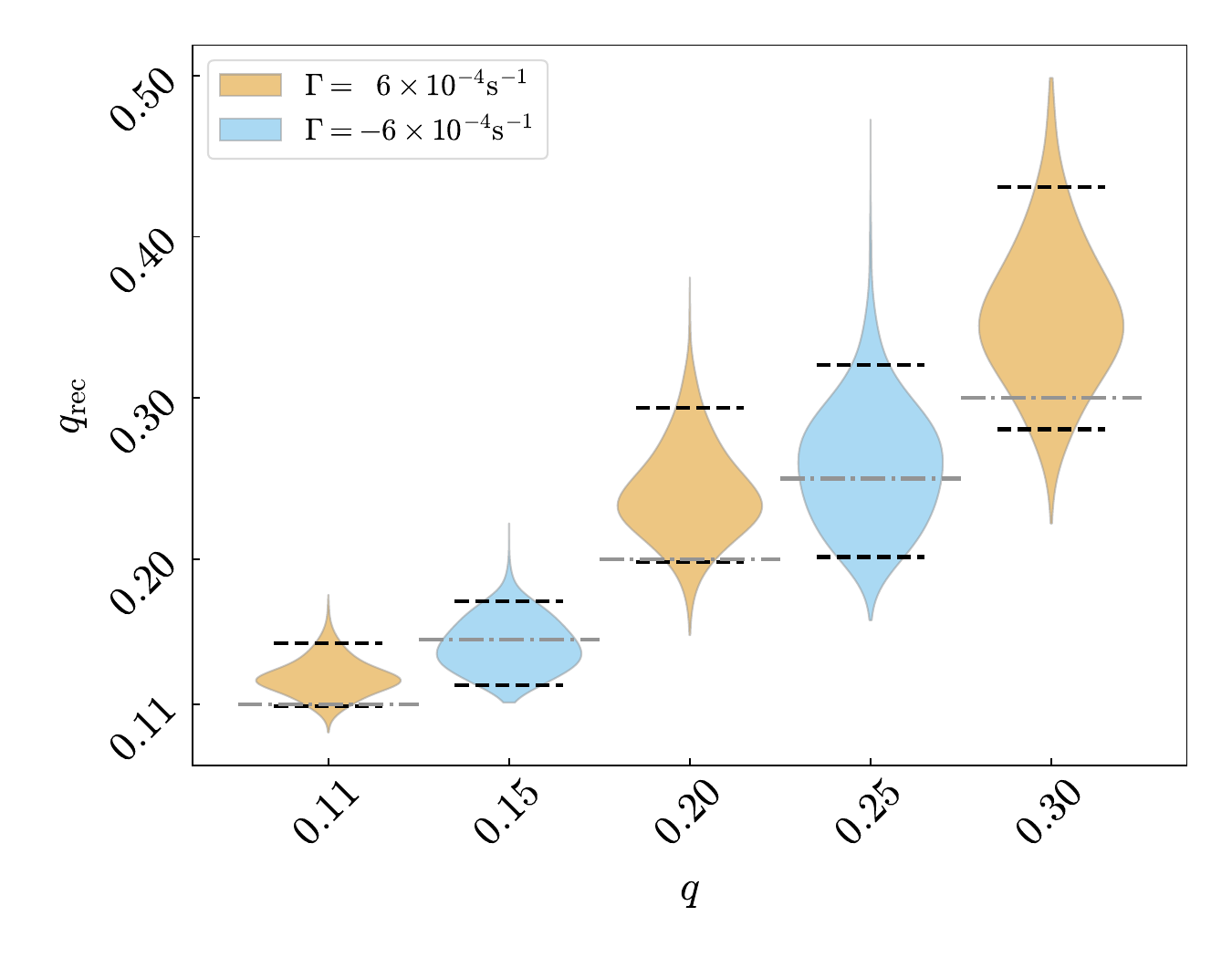}
     \includegraphics[width=0.485\linewidth]{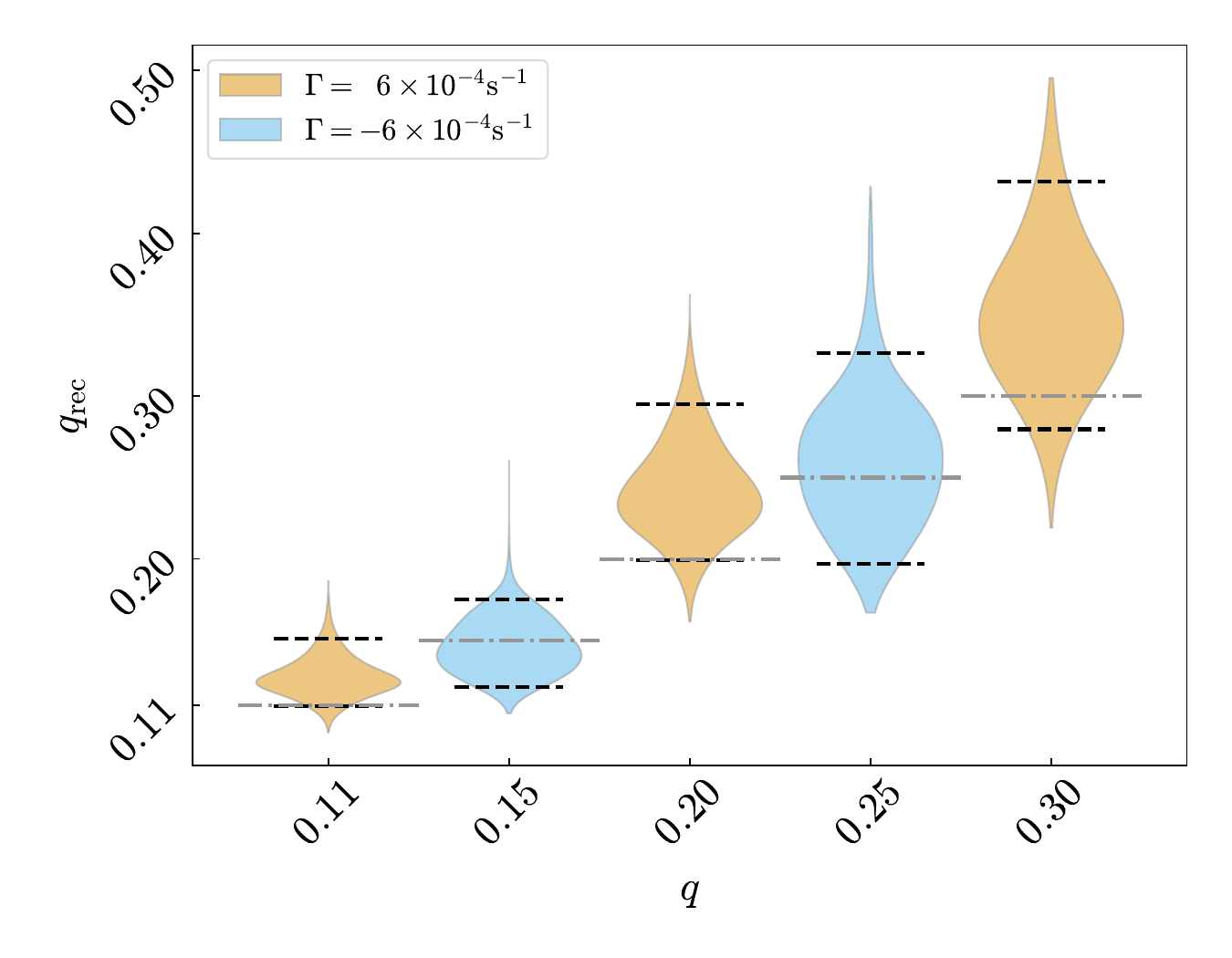}     
    \caption{The \textit{left panel} represents the posteriors of $\mathcal{M}$ ({\it top panel}) and $\Gamma$ ({\it bottom panel}) when the recovery waveform has (2,2) mode LOSA correction to all subdominant modes, while the \textit{right panel} shows the posteriors of the same when the recovery waveform has (2,2) mode LOSA correction only to the (2,2) mode, while the subdominant modes remain unaffected. The dashed lines represent the 90\% CI, while the dash-dot lines represent the injected values.}
    \label{fig: bias_O4}
\end{figure*}

Figure~\ref{fig: bias_O4} shows the posteriors of $\mathcal{M}$ and $\Gamma$ for different mass ratios for O4 detector sensitivities. The \textit{left} panel in both plots corresponds to the bias resulting from the implementation of the corrections following Section~\ref{sec: all_modes_bias}, and the \textit{right} panel corresponds to the same bias resulting from the implementation of the corrections following Section~\ref{sec: only_22_mode_bias}. The blue posteriors correspond to $\Gamma = -6 \times 10^{-4} {\rm s^{-1}}$, while the orange ones correspond to $\Gamma = 6 \times 10^{-4} {\rm s^{-1}}$. Similar to the O5 case, here as well, $q$ is recovered within the 90\% CI, and hence we do not show its posterior. We find that $\Gamma$ remains biased for $q < 0.25$ when the injected LOSA is negative and for $q \sim 0.3$ when the injected LOSA is positive in both scenarios. For $\mathcal{M}$, we find that $\mathcal{M}$ remains biased till $q \sim 0.15$ for negative LOSA injection and till $q \sim 0.3$ for positive LOSA injection. Notice that in O4, even for high $\Gamma$, the bias disappears at relatively lower mass ratios.

\section{Case Study : GW190814}\label{app: gw_pe}
\begin{figure*}[ht!]
    \centering
    \includegraphics[width=0.485\linewidth]{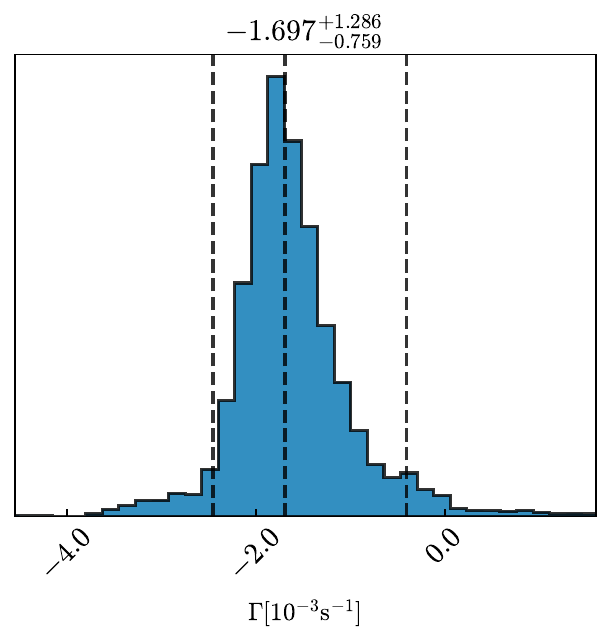}
     
    \caption{The posterior of $\Gamma$ for GW190814. Though zero is excluded from the 90\% CI, the non-accelerating quasi-circular model is preferred over the accelerating one with $\ln (\rm BF) = 1.04$.}
    \label{fig: real_event}
\end{figure*}

We apply the procedure delineated in this paper to add LOSA corrections to higher harmonics on GW190814. Figure~\ref{fig: real_event} shows the posterior on $\Gamma$ for the same. We find that it agrees with~\cite{Roy:2026mco, Pompili:2026tdf} and the 90\% CI does not accommodate \textit{zero}. However, a Bayes factor (BF) check reveals that the $\ln(\rm BF)$ of the vacuum hypothesis with respect to the LOSA hypothesis is 1.04, suggesting moderate evidence for a \textit{zero} LOSA detection. Therefore, one cannot claim a LOSA detection.

\end{document}